\documentclass[pra,twocolumn,floatfix,a4paper,superscriptaddress,longbibliography]{revtex4-1}
\usepackage{bbm,color,graphicx,amsmath,txfonts}
\usepackage{hyperref}

\newcommand{\rp}[1]{(\ref{#1})}

\newcommand{\abs}[1]{\left|{#1}\right|}

\newcommand{\rabs}[1]{\left.{#1}\right|}

\newcommand{\av}[1]{\left\langle #1 \right\rangle}

\newcommand{\wt}[0]{\widetilde}
\newcommand{\oo}[0]{^{\circ}}
\newcommand{\al}[1]{^{(#1)}}
\newcommand{\da}{^\dagger}

\newcommand{\pt}[1]{\left( #1 \right)}
\newcommand{\pq}[1]{\left[ #1 \right]}
\newcommand{\pg}[1]{\left\{ #1 \right\}}

\newcommand{\lpq}[1]{\left[ #1 \right.}
\newcommand{\lpg}[1]{\left\{ #1 \right.}

\newcommand{\rpq}[1]{\left. #1 \right]}
\newcommand{\rpg}[1]{\left. #1 \right\}}
\newcommand{\ee}{{\rm e}}
\newcommand{\ii}{{\rm i}}
\newcommand{\dd}{{\rm d}}

\newcommand{\id}{\mathbbm{1}}

\newcommand{\nn}{{\nonumber}}

\newcommand{\pp}[2]{ {\mbox{\scriptsize$
                      \begin{array}{#1}                       
                       #2
                      \end{array}$} }    }

\newcommand{\ovl}{\overline}

\newcommand{\va}{{\bf a}}

\newcommand{\EE}{{\cal E}}

\newcommand{\SSS}{{\cal S}}

\definecolor{blue}{rgb}{0,0,0.8}

\definecolor{green}{rgb}{0,0.6,0}

\usepackage[normalem]{ulem}
\newcommand{\stkout}[1]{\ifmmode\textrm{\sout{\ensuremath{#1}}}\else\sout{#1}\fi}

\begin{document}

\title{Optics-assisted enhanced sensing at radio frequencies in an optoelectromechanical system}

\author{Najmeh Eshaqi-Sani}
\email{naj.eshaqi@gmail.com}
\affiliation{Physics Division, School of Science and Technology, University of Camerino, I-62032 Camerino (MC), Italy}
\affiliation{School of Physics, Institute for Research in Fundamental Sciences (IPM), P.O. Box 19395-5531, Tehran, Iran}

\author{Stefano Zippilli}
\email{stefano.zippilli@unicam.it}
\affiliation{Physics Division, School of Science and Technology, University of Camerino, I-62032 Camerino (MC), Italy}

\author{David Vitali}
\email{david.vitali@unicam.it}
\affiliation{Physics Division, School of Science and Technology, University of Camerino, I-62032 Camerino (MC), Italy}
\affiliation{INFN, Sezione di Perugia, via A. Pascoli, Perugia, Italy}
\affiliation{CNR-INO, L.go Enrico Fermi 6, I-50125 Firenze, Italy}

\begin{abstract}

We investigate a scheme to enhance the sensitivity in detecting weak variations in a parameter of an optoelectromechanical system by detecting the system response at radio frequencies. 
We consider a setup, where either one or two mechanical modes mediate the interaction between an optical cavity and an rf resonator. 
This system can be operated in a regime of impedance matching where thermal fluctuations are redistributed among the system elements, and, in particular, rf output noise can be reduced to the quantum vacuum noise level.
We show that this effect can be used to boost the sensitivity in detecting parameter variations also in regimes of high thermal noise. We characterize the performance of this protocol in detecting variations in the capacitance of the rf resonator.
\end{abstract}

\date{\today}
\maketitle

\section{Introduction}

The quest for ever better and more sensitive detection strategies is central in physics. Quantum sensing refers to a variety of strategies that attempt to improve the ability to detect small signals exploiting quantum effects, processes or systems~\cite{Degen,pirandola2018}. In this context the precise
control and sensitive detection of rf signals are essential for a wide range of modern technologies, including communication, sensing, and highly sensitive astronomical metrology. However, thermal noise represents a fundamental challenge for quantum operations in this frequency  range~\cite{bagci2014,moaddelhaghighi2018,takeda2018b,simonsen2019b,Malossi,Jiang,Bonaldi}. 
It is, therefore, interesting to study strategies to mitigate the detrimental effects of noise at rf frequencies.
Various works~\cite{barzanjeh2018,Eshaqi,MetelmannEnt,tang2023} discuss the possibility to manipulate and reroute thermal noise in multimode systems. Here, we employ a similar redistribution of thermal fluctuations in an optoelectromechanical system to reduce the noise on the measurement of a MHz radio frequency (rf) fieldand thus to enhance the corresponding signal-to-noise ratio.

In particular, we consider an optoelectromechanical system where one or two vibrational modes of a mechanical object mediate the interaction between the light of an optical cavity and the rf field of an LC resonator. 
We focus on a regime of impedance matching where an rf input field experiences no reflection. In this regime, the noise of the fields is redistributed among the elements of the system, such that noise in the rf output can be strongly reduced approaching the quantum vacuum noise level. Here, we show that an rf probe field reflected from this system exhibits reduced noise. Consequently, detecting variations in a system parameter using this probe field may show significantly enhanced sensitivity compared with a similar measurement conducted with a simple rf resonator. Notably, this enhancement is observed over a wide range of temperatures, and it is effective also with noisy rf systems operating in the high-temperature regime. Here we analyze this effect to probe small perturbations in the capacitance of the rf resonator. 

We analyze and compare the performance of this system when either one or two mechanical resonators mediate the interaction between optical and rf photons. In the latter case, we also analyze whether the regime of nonreciprocity, as demonstrated  in Ref.~\cite{Eshaqi}, can be instrumental for sensing. In fact, nonreciprocity in optomechanical systems~\cite{Clerk1,Xu,Tian,Bernier,Malz,Barzanjeh,Eshaqi}
has been suggested as a potential tool to achieve enhanced sensing~\cite{Clerk,mcdonald2020,Nori}. Therefore, it is interesting to ask whether nonreciprocity can be exploited to further enhance the detection sensitivity. We demonstrate that in our case, nonreciprocity does not provide any specific advantage.

The paper is organized as follows. In Sec.~\ref{model} we describe the system and we discuss how to model the perturbed capacitance that we aim to probe.
Then, in Sec.~\ref{sensing} we introduce the detection strategy and we compute the signal-to-noise ratio achievable in the nonreciprocal regime. In Sec.~\ref{results} we discuss the performance of our protocol and compare it with the performance of a similar detection protocol realized with a simple rf circuit. Finally, in Sec.~\ref{conclusions} we draw our conclusions. In the appendices we present additional details about our analytical description, the maximization of the signal-to-noise ratio, and the detection noise and efficiency in our model.

\begin{figure}[t!]
\includegraphics[angle=0, width=1\linewidth]{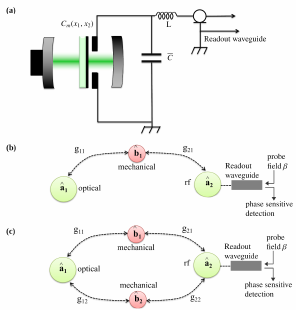}
\caption{
(a) Setup of the proposed optoelectromechanical system, where the capacitance $C$ introduced in the text corresponds to the sum
$C=C_m+\ovl C$ of the two capacitances $C_m$ and $\ovl C$ depicted in this figure. 
(b) and (c), schematic representation of the system with a single (b) and two (c) vibrational modes. In the text we will refer to these two situations as the 3-modes and 4-modes models respectively.
The transmission from the optical input to the rf output or, vice versa, from the rf input to the optical output are mediated by the 
mechanical resonators. $\beta$ is the amplitude of the probe field applied to the readout waveguide coupled to the mode $a_{2}$.
}
\label{fig1}
\end{figure}

\begin{figure}[t!]
\includegraphics[angle=0, width=0.8\linewidth]{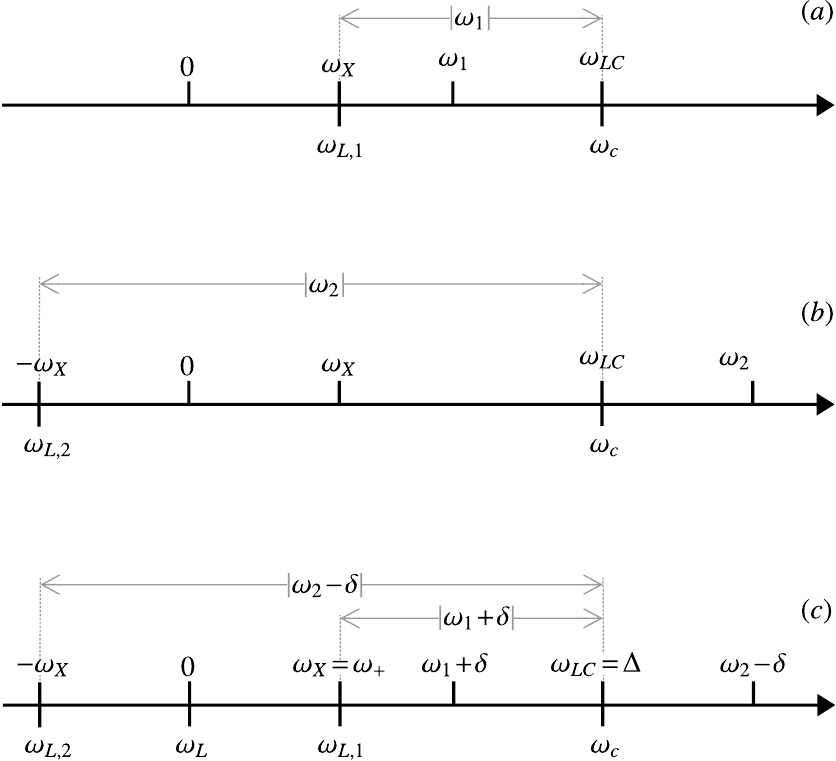}
\caption{
Frequency scheme for the 3-modes [(a) and (b)] and 4-modes [(c)] models (see Fig.~\ref{fig1}). In each panel, the optical scale is  below the horizontal axis and the radio-frequency scale above.  
In (a) and (b) the mechanical frequency is larger and smaller then the rf-frequency, respectively. 
}
\label{fig1.2}
\end{figure}

%%%%%%%%%%%%%%%%%%%%%%%%%%%%%%%%%%%%%%%%%%%%%%%%%%%%%%%%%%%%%%
\section{The system}
\label{model}

We study a hybrid optoelectromechanical system where an optical cavity is coupled by radiation pressure to either one or two resonant vibrational modes of a mechanical element and which, in turn, is/are capacitively coupled to an rf resonant LC circuit.  
We will refer to the situation with one mechanical mode as the 3-modes model, and that with two mechanical modes as the 4-modes model (see Fig.~\ref{fig1}).
The 4-modes model,
is similar to the one analyzed in Ref.~\cite{Eshaqi}. Moreover, having the 4-modes model, the 3-modes one is immediately found by setting the corresponding coupling strengths $g_{j2}$  to zero [see Fig.~\ref{fig1} (b) and (c)]. 

In this section, we introduce the more general 4-modes model of Ref.~\cite{Eshaqi}, with two mechanical resonators, which we will adapt as needed to the case of a single mechanical resonator (3-modes model) in the following sections.
Differently from Ref.~\cite{Eshaqi}, here we assume a small perturbation affecting the capacitance, and our objective is to measure this perturbation.

We assume systems operated in the resolved sideband regime with mechanical frequencies $\omega_j$ much larger than the dissipation rates of the electromagnetic modes $\omega_j\gg \kappa,\gamma_{LC}$, where $\kappa$  and $
\gamma_{LC}$ are, respectively, the optical and rf decay rates.
We also assume driving fields near-resonant to the red mechanical sidebands (see Fig.~\ref{fig1.2}) and linearized electro-/opto-mechanical coupling strength sufficiently small such that 
one can perform a rotating wave approximation, whereby the resulting Hamiltonian interaction terms conserve the number of excitations. 
Note that, when considering a rf-mode at frequency $\omega_{LC}$ of the same order of $\omega_j$,
the red-sideband driving can be obtained with a single driving field (at frequency $\omega_X$) also in the case of two mechanical modes~\cite{Eshaqi}, see Fig.~\ref{fig1.2} (c).

\subsection{The system Hamiltonian}

As in Ref.~\cite{Eshaqi}, the system Hamiltonian is given by the sum of an optical, mechanical and electrical contribution according to 
\begin{equation}
  \hat{H}=\hat{H}_{\rm opt}+\hat{H}_{\rm mech}+\hat{H}_{\rm LC}
  \label{hin}
\end{equation}
where
the Hamiltonian for the two mechanical modes with displacements $\hat x_j$ and frequencies $\omega_j$ is
\begin{equation}
\hat{H}_{\rm mech} = \sum_{j=1,2}\frac{\hat{p}_{j}^2}{2 m_{j}}+\frac{m_{j} \omega_{j}^{2} \hat{x}_{j}^2}{2}\ ,
\end{equation}
the optical term that describes the optical cavity at frequency $\omega_c(\hat x_1,\hat x_2)$ which depends on the mechanical displacement, and includes the two-tone driving at frequencies $\omega_{Lj}$ and strength $\EE_j$ (for $j\in\pg{1,2}$), is given by
\begin{eqnarray}
\hat{H}_{\rm opt}&=& \hbar \omega_{c}({x}_{1},{x}_{2})\:\hat{a}_{1}^\dag \hat{a}_{1} 
+\hbar[(\EE_{1} e^{-\, i\,\omega_{L1}t}+\EE_{2} 
e^{-\, i\,\omega_{L2}t})\hat{a}_{1}^\dag 
\nn\\
&+&
\hbar[(\EE_{1}^* e^{i\,\omega_{L1}t}+\EE_{2}^* 
e^{i\,\omega_{L2}t})\hat{a}_{1} 
] ,
\end{eqnarray}
and finally the contribution of the rf-circuit, with a capacitance $C(\hat x_1,\hat x_2)$ which depends on the mechanical displacements, includes the driving, with strength $V_{AC}$, and at frequency $\omega_X$, such that
\begin{equation}
\hat{H}_{\rm LC} = \frac{\hat{\phi}^2}{2 L}+\frac{\hat{q}^2}{2 {C}(\hat{x}_{1},\hat{x}_{2})}- \hat{q}V_{\rm AC}\cos(\omega_{X}t-\phi_{X}) \ .
\end{equation} 

In particular, the total Hamiltonian can be expressed in terms of the annihilation and creation operators for the mechanical excitations and for the rf-photons ($\hat b_j$, $\hat b_j\da$, $\hat a_2$ and $\hat a_2\da$), such that 
\begin{equation}
\hat{x}_{j}\equiv x_{zpf,j}\,
\pt{\hat{b}_{j}+\hat{b}_{j}^{\dagger}}\ ,
\end{equation}
\begin{equation}
\hat{p}_{j}\equiv p_{zpf,j}\frac{\hat{b}_{j}-\hat{b}_{j}^{\dagger}}{i}\ ,
\end{equation}
with $x_{zpf,j}\equiv \sqrt{\frac{\hbar}{2m_{j}\omega_{j}}}$ and $p_{zpf,j}\equiv m\omega_{j} x_{zpf,j}$, 
and 
\begin{equation}
\hat{q}\equiv \bar{q}_{zpf}\,
\pt{\hat{a}_{2}+\hat{a}_{2}^{\dagger}}\ ,
\end{equation}
\begin{equation}
\hat{\phi}\equiv \bar{\phi}_{zpf}\frac{\hat{a}_{2}-\hat{a}_{2}^{\dagger}}{i}\ ,
\end{equation}
with $ q_{zpf}\equiv \sqrt{\frac{\hbar}{2\,L\,\omega_{LC}^{(0)}}}$ and $\phi_{zpf}\equiv \sqrt{\frac{L\,\hbar\omega_{LC}^{(0)}}{2}}$, where $\omega_{LC}^{(0)}={1}/{\sqrt{L\,C_0}}$, with 
\begin{eqnarray}
C_0\equiv C(0,0)\ ,
\end{eqnarray}
is the resonant frequency of the rf-circuit.
For small mechanical displacements, this Hamiltonian can be expanded at first order in $\hat x_j$, and can be expressed as follows, in the reference frame, for the optical mode, rotating at the frequency $(\omega_{L1}+\omega_{L2})/2$, (see Ref. \cite{Eshaqi})
\begin{eqnarray} \label{H}
\hat{H}&=& \hbar \Delta_{L}\hat{a}_{1}^\dag \hat{a}_{1}+\hbar\sum_{j=1,2}g_{0,1j}(\hat{b}_{j}+\hat{b}_{j}^{\dagger})\hat{a}_{1}^{\dagger}\hat{a}_{1}
\nn\\\nonumber
&+&\sum_{j=1,2}\hbar\omega_{j}\hat{b}_{j}^{\dagger} \hat{b}_{j} + \hbar\omega_{LC}^{(0)}\hat{a}_{2}^{\dagger} \hat{a}_{2}
\\\nonumber
&-&\hbar\sum_{j=1,2}g_{0,2j}(\hat{b}_{j}+\hat{b}^{\dagger}_{j})(\hat{a}_{2}+\hat{a}_{2}^{\dagger})^{2}
\\\nonumber
&+& \hbar[(\mathcal{E}_{1} e^{-i\omega_{+}t}+\mathcal{E}_{2} e^{
i\omega_{+}t})\hat{a}_{1}^\dag +h.c.]
\nn\\&-&
\hbar\pt{{V'}^*\,e^{i\omega_{X}t}+V'\,e^{-i\omega_{X}t}}\pt{\hat a_2+\hat a_2\da}\ ,
\end{eqnarray}
where $\Delta_{L}= \omega_{c}(0,0)-(\omega_{L1}+\omega_{L2})/2$ is the  optical detuning,
$\omega_+=(\omega_{L1}-\omega_{L2})/2$, $V'=e^{i\phi_{X}}\ q_{zpf}V_{\rm AC}/2\hbar$ the effective rf driving strength, and finally,
$g_{0,1j}\equiv \frac{\partial\omega_{c}}{\partial x_{j}}|_{x_{i}=0}x_{zpf,j}$ and $g_{0,2j}\equiv \frac{\omega_{LC}^{(0)}}{4 C_0}x_{zpf,j}\,C'_j$
with 
\begin{eqnarray}
C'_j\equiv\frac{\partial C}{\partial x_{j}}\Bigl|_{x_{j}=0}\ ,
\end{eqnarray}
are the bare coupling strengths between the mechanical modes and the electromagnetic fields. 

As usual in optomechanics the relevant degrees of freedom are the fluctuations about the corresponding field averages $\alpha_j(t)$ and $\beta_j(t)$,	
\begin{eqnarray}\label{deltaa-}
\delta\hat{a}_{j}&=&\hat{a}_{j}-\alpha_{j}(t)
\\
\delta\hat{b}_{j}&=&\hat{b}_{j}-\beta_{j}(t)\ .
\label{deltab}
\end{eqnarray}
As shown in detail in Ref.~\cite{Eshaqi}, for sufficiently strong driving fields, the dynamics of the fluctuations can be linearized about their average fields. 
In this limit the electromagnetic field frequencies are shifted by a quantity proportional to the amplitude of the mechanical oscillations. Specifically, the optical detuning and the rf frequency acquire the values
$\Delta\equiv\Delta_{L}+2\sum_{j} g_{0,1j} Re\pg{\beta_{j}\al{dc}}$
and  $\omega_{LC}\equiv \omega_{LC}^{(0)}-4\sum_{j} g_{0,2j} Re\pg{\beta_{j}\al{dc}}$,
with $\beta_j\al{dc}$ the time-independent part of the mean mechanical amplitude $\beta_j(t)$~\cite{Eshaqi}.
Correspondingly the linearized Hamiltonian for the fluctuations, expressed in interaction picture with respect to
\begin{eqnarray}\label{H0}
\hat{H}_{0}&=&
\hbar\Delta\ \delta\hat{a}_{1}^{\dagger}\,\delta\hat{a}_{1}+\hbar\omega_{LC}\,\delta\hat{a}_{2}^{\dagger}\,\delta\hat{a}_{2}+\hbar\pt{\omega_{1}+\delta}\,\delta\hat{b}_{1}^{\dagger}\,\delta\hat{b}_{1}
\nn\\&&
+\hbar\pt{\omega_{2}-\delta}\,\delta\hat{b}_{2}^{\dagger}\,\delta\hat{b}_{2}\ ,
\end{eqnarray} 
is given by~\cite{Eshaqi}
\begin{eqnarray}\label{Heff}
\hat{H}_{eff}&=&-\hbar~\delta~ \delta\hat{b}_{1}^{\dagger}\delta\hat{b}_{1}+ \hbar~\delta ~\delta\hat{b}_{2}^{\dagger}\delta\hat{b}_{2}
\nn\\
&+&\hbar(g_{11}\delta\hat{a}_{1}^{\dagger}\delta\hat{b}_{1}
+g_{11}^{*}\delta\hat{a}_{1}\delta\hat{b}_{1}^{\dagger})
\nn\\
&+&\hbar(g_{12}\delta\hat{a}_{1}^{\dagger}\delta\hat{b}_{2}
+g_{12}^{*}\delta\hat{a}_{1}\delta\hat{b}_{2}^{\dagger})
\nn\\
&-&\hbar(g_{21}\delta\hat{a}_{2}\delta\hat{b}_{1}^{\dagger}
+g_{21}^{*}\delta\hat{a}_{2}^{\dagger}\delta\hat{b}_{1})
\\\nonumber
&-&\hbar(g_{22}\delta\hat{a}_{2}\delta\hat{b}_{2}^{\dagger}
+g_{22}^{*}\delta\hat{a}_{2}^{\dagger}\delta\hat{b}_{2})
\end{eqnarray}
where
\begin{eqnarray}\label{GGGG}
g_{11}&=&-i\frac{\ g_{0,11}\ \EE_{1}
}{\kappa+i(\Delta_L-\omega_+)}
\nn\\
g_{12}&=&-i\frac{\ g_{0,12}\ \EE_{2}
}{\kappa+i(\Delta_L+\omega_+)}
\nn\\
g_{21}&=&2\,i\ g_{0,21}\ 
\pg{\frac{1}{\frac{\gamma_{LC}}{2}+i(\omega_{LC}^{(0)}+\omega_X)}
-\frac{1}{\frac{\gamma_{LC}}{2}-i(\omega_{LC}^{(0)}-\omega_X)}
}\, {V'}^*
\nn\\
g_{22}&=&2\,i\ g_{0,22}\ 
\pg{\frac{1}{\frac{\gamma_{LC}}{2}+i(\omega_{LC}^{(0)}-\omega_X)}
-\frac{1}{\frac{\gamma_{LC}}{2}-i(\omega_{LC}^{(0)}+\omega_X)}
}\, {V'}\ ,
\nn\\
\end{eqnarray}
with $\gamma_{m,j}$, $\kappa$ and $\gamma_{LC}$ the dissipation rates of, respectively, the $j$-th mechanical resonator, the optical cavity and the rf-circuit.
Moreover, the frequencies fulfill the following resonant conditions 
\begin{eqnarray}\label{resonance}
\Delta-\omega_+&=&\omega_{LC}-\omega_X=\omega_1+\delta
\nn\\
\Delta+\omega_+&=&\omega_{LC}+\omega_X=\omega_2-\delta\ ,
\end{eqnarray}
such that the system is driven on (or close to) the red mechanical sidebands,  with $\delta$ being a small detuning ($\abs{\delta}\lesssim \abs{g_{\ell j}}$) from the exact red sideband driving condition. 
This Hamiltonian is valid if the optomechanical couplings $g_{\ell j}$ are sufficiently small to allow for a rotating wave approximation under which one can neglect non-resonant processes, and if the bare couplings $g_{0,\ell j}$ are sufficiently small to allow for a perturbative evaluation of the average fields at the lowest relevant order (see Ref.~\cite{Eshaqi} for further details). 

These equations can be easily adapted to the case with a single mechanical resonator by setting either $g_{0,j2}=0$ or $g_{0,j1}=0$ for $j=1,2$. In this case, the resonant conditions, analogous to Eq.~\rp{resonance},
are obtained by detuning the frequency of the optical driving field 
by an amount equal to the mechanical frequency, 
either $\omega_c-\omega_{L1}=\omega_1$ or $\omega_c-\omega_{L2}=\omega_2$, from the resonance of the optical cavity. Similarly, the rf driving frequency has to fulfill either the relation
$\omega_{LC}-\omega_X=\omega_1$ or $\omega_{LC}+\omega_X=\omega_2$ (see Fig.~\ref{fig1.2}). If the conditions for the rotating wave approximations, similar to the ones employed in the previous case~\cite{Eshaqi}, are fulfilled, the system Hamiltonian reduces to a form equal to Eq.~\rp{Heff} with either $g_{j2}=0$ or $g_{j1}=0$. In this 
3-modes
case we consider only the situation with $\delta=0$.

\subsection{The perturbed capacitance and its effect on the system Hamiltonian}

Differently from  Ref.~\cite{Eshaqi}, here, we assume that the capacitance of the rf-resonator is influenced by external factors, causing a small perturbation. Here we discuss how to model the perturbation in our system, then, in the following section we introduce the scheme for the detection of this perturbation. 

We incorporate the perturbation's effect into the earlier equations by adjusting the capacitance using 
a small parameter $\epsilon\ll 1$ according to
\begin{eqnarray}
\label{pert}
C(x_1,x_2)\ \to\ C(x_1,x_2)(1+\epsilon)
\end{eqnarray}
(note that we assume that the perturbation doesn't depend on the mechanical displacements).
In particular, we approximate the previous Hamiltonians by retaining only linear terms in $\epsilon$. Using the substitutions 
$C_0\ \to\ C_0(1+\epsilon)$ and $C'_j\ \to\ C'_j(1+\epsilon)$, we find
\begin{eqnarray}
{\omega}_{LC}^{(0)}
\Bigl|_{C_0\ \to\ C_0(1+\epsilon)}
&\simeq& \omega_{LC}^{(0)}(1{-\frac{\epsilon}{2}})
\nn\\
{q}_{zpf}\Bigl|_{C_0\ \to\ C_0(1+\epsilon)}
&\simeq & q_{zpf}(1+\frac{\epsilon}{4})
\nn\\
{V}^{\prime}\Bigl|_{C_0\ \to\ C_0(1+\epsilon)}
&\simeq& V^{\prime} (1+\frac{\epsilon}{4})
\nn\\
{g}_{o,2j}\Bigl|_{\pp{l}
{
{C_0\ \to\ C_0(1+\epsilon)}
\\
C'_j\ \to\ C'_j(1+\epsilon)
}}
&\simeq & g_{0,2j}(1-\frac{\epsilon}{2})\ .
\end{eqnarray}
Thereby, the total system Hamiltonian corresponding to Eq.~\rp{H}, including the perturbation, can be expressed as follows
\begin{eqnarray}    
\hat H\al{\epsilon}
=\hat H\Bigl|_{\pp{l}
{
{C_0\ \to\ C_0(1+\epsilon)}
\\
C'_j\ \to\ C'_j(1+\epsilon)
}}
\simeq 
\hat H+\epsilon\ \hat W
\end{eqnarray}
where $\hat W$ describes the effect of the perturbation and is given by
\begin{eqnarray}
\hat W&=&\frac{\hbar}{2}\sum_{j=1,2} g_{0,2j}(\hat{a}_{2}+\hat{a}_{2}^\dag)^{2}(\hat{b}_{j} +\hat{b}_{j}^\dag)
-\frac{\hbar}{2}\omega_{LC}^{(0)}\:\hat{a}_{2}^\dag \hat{a}_{2}
\nn\\&&-
\frac{\hbar}{4}\pt{{V'}^*\,e^{i\omega_{X}t}+V'\,e^{-i\omega_{X}t}}\pt{\hat a_2+\hat a_2\da}
\ .
\end{eqnarray} 
This results indicates that the perturbed capacitance induces small modifications of the rf driving, of the rf frequency and of the coupling strength between mechanical modes and rf-circuit.

Correspondingly, we find 
\begin{eqnarray}\label{gepsilon}
g_{2j}\Bigl|_{\pp{l}
{
{C_0\ \to\ C_0(1+\epsilon)}
\\
C'_j\ \to\ C'_j(1+\epsilon)
}}&\simeq&
g_{2j} + \epsilon\ \wt g_{2j}
\\
\label{omegaLCepsilon}
\omega_{LC}\Bigl|_{\pp{l}
{
{C_0\ \to\ C_0(1+\epsilon)}
\\
C'_j\ \to\ C'_j(1+\epsilon)
}}&\simeq&
\omega_{LC}\pt{1-\frac{\epsilon}{2}}-\epsilon\ \wt\omega_{LC}\ ,
\\
\Delta\Bigl|_{\pp{l}
{
{C_0\ \to\ C_0(1+\epsilon)}
\\
C'_j\ \to\ C'_j(1+\epsilon)
}}&\simeq&\Delta+\epsilon\ \wt\Delta\ ,
\label{Deltaepsilon}
\end{eqnarray}
where the parameters $\wt\omega_{LC}$, $\wt\Delta$ and  $\abs{\wt g_{2j}}$ (the explicit form of which is reported in App.~\ref{wtpars}) are small
\begin{eqnarray}
\omega_{LC}\gg\wt\omega_{LC},\wt\Delta,\abs{\wt g_{2j}}\ ,
\end{eqnarray}	
(in following sections we will show that their effect can be neglected)
so that the dominant effect of the perturbation is to induce a shift of the LC frequency of magnitude $\sim\ \epsilon\,\omega_{LC}/2$.  
As a result, the Hamiltonian for the fluctuations, corresponding to Eq.~\rp{Heff}, is given by
\begin{eqnarray}    
\hat H_{eff}\al{\epsilon}
=\hat H_{eff}\Bigl|_{\pp{l}
{
{C_0\ \to\ C_0(1+\epsilon)}
\\
C'_j\ \to\ C'_j(1+\epsilon)
}}
\simeq 
\hat H_{eff}+\epsilon\ \hat W_{eff}
\end{eqnarray}
with
\begin{eqnarray}
\hat W_{eff}&=&
\hbar\ \wt\Delta\ \delta\hat{a}_{1}^{\dagger} \delta\hat{a}_{1}
-\hbar\ \pt{\frac{\omega_{LC}}{2}+\wt\omega_{LC}}\,\delta\hat{a}_{2}^{\dagger} \delta\hat{a}_{2}
\nn\\&&
-\hbar(\wt g_{21}\delta\hat{a}_{2}\delta\hat{b}_{1}^{\dagger}
+\wt g_{21}^{*}\delta\hat{a}_{2}^{\dagger}\delta\hat{b}_{1})
\nn\\&&
-\hbar(\wt g_{22}\delta\hat{a}_{2}\delta\hat{b}_{2}^{\dagger}
+g_{22}^{*}\delta\hat{a}_{2}^{\dagger}\delta\hat{b}_{2}) \ .
\end{eqnarray}

\section{Sensing} 
\label{sensing}	
	
In order to detect the perturbation, we make use of a weak coherent rf probe field and measure the reflected rf signal with a phase sensitive detection strategy (see App.~\ref{App:detection}) following the general approach detailed in Ref.~\cite{Clerk}.
	
\subsection{The quantum Langevin equations}	
	
The dissipative dynamics of our system can be described in terms of quantum Langevin equations for the fluctuations 
\begin{eqnarray}\label{QLEfig}
\dot{\delta\hat{a}}_{1} &=& -\pt{\kappa+i\,\wt\Delta}\,\delta\hat a_1
-i\,\sum_{j=1,2}\,g_{1j}\ \delta\hat{b}_{j}
+ \sqrt{2 \kappa}\,\hat{a}_{1}^{(in)}
\nn\\
\dot{\delta\hat{a}}_{2} &=& -\pq{
\frac{\gamma_{LC}}{2}-i
\epsilon\,
\pt{\frac{\omega_{LC}}{2}+\wt\omega_{LC}}
}\,\delta\hat{a}_{2}
\nn\\&&
+i\,\sum_{j=1,2}\,\pt{{g}_{2j}^*+\epsilon\,\wt{g}_{2j}^*}\ \delta\hat{b}_{j}
\nn\\&&
+\sqrt{\gamma_{LC}}\,\hat{a}_{2}^{(in)}\nn\\
\dot{\delta\hat{b}}_{j} &=& -\pq{\frac{\gamma_{m,j}}{2}+i(-1)^j\,\delta}\,\delta\hat{b}_{j}
-i\,g_{1 j}^*\,\delta\hat{a}_1
\nn\\&&
+i\,\pt{{g}_{2 j}+\epsilon\,\wt{g}_{2 j}}
\,\delta\hat{a}_2
\nn\\&&
+ \sqrt{\gamma_{m,j}}\,\hat{b}_{j}^{(in)}\ .
\end{eqnarray}
where the input noise operators are uncorrelated and characterized by thermal correlations at temperature $T$
$\langle \hat{b}_{j,in}^{\dagger}(t)\hat{b}_{j,in}(t')\rangle=
\bar{n}_{b\,j}\ \delta(t-t')$, and $\langle \hat{a}_{j,in}^{\dagger}(t)\hat{a}_{j,in}(t')\rangle= \tilde{n}_{a\,j}\ \delta(t-t')$, 
with the number of thermal phonons given by $\bar{n}_{b\,j}=\left\{\exp[\hbar\omega_{j}/k_B T]-1\right\}^{-1}$, $j=1,2$, the number of thermal rf photons is $\tilde{n}_{a\,2}=\left\{\exp[\hbar\omega_{LC}^{(0)}/k_B T]-1\right\}^{-1}$, and $\tilde{n}_{a\,1}\simeq 0$.

To detect the small perturbation, we assume that we can inject a coherent field at frequency $\omega_{LC}$ and with amplitude $\beta$ through a readout waveguide (see Fig.~\ref{fig1}) coupled to the rf resonator.
This can be modeled by adding a coherent component to the input noise operator of the rf-circuit, such that
\begin{eqnarray}\label{beta}
\hat a_2\al{in}\ \to\ \hat a_2\al{in} + \beta\ .
\end{eqnarray}
In particular, we assume that $\beta$ is sufficiently weak such that the derivation of the approximated model discussed in the previous sections remains valid. 

It is now useful to express the quantum Langevin equation in matrix form. We consider the vectors of operators $\va=\pt{\delta\hat a_1,\delta\hat a_2,\delta\hat b_1,\delta\hat b_2}^T$ and $\va_{in}=\pt{\hat a_1\al{in},\hat a_2\al{in}+\beta,\hat b_1\al{in},\hat b_2\al{in}}^T$, so that Eq.~\rp{QLEfig} can be expressed as
\begin{eqnarray}
\dot\va=\pt{M+\epsilon V}\va+L\ \va_{in}
\end{eqnarray}
where the matrices of coefficients are
\begin{eqnarray}
M=-\left(\begin{array}{cccc}
\kappa & 0 & i\,g_{11} & i\,g_{12}
\\
0 & \frac{\gamma_{LC}}{2} & -i\,g_{21}^{*} & -i\,g_{22}^{*}
\\
i\,g_{11}^{*} & -i\,g_{21} & \frac{\gamma_{m,1}}{2}-i\delta & 0
\\
i\,g_{12}^{*}& -i\,g_{22} & 0 & \frac{\gamma_{m,2}}{2}+i\delta
\end{array}\right) ,
\end{eqnarray}
\begin{eqnarray}\label{VVV}
V=i\,
\left(
\begin{array}{cccc}
-\wt\Delta & 0 & 0 & 0
\\
0 & \frac{\omega_{LC}}{2}+\wt\omega_{LC} & \wt g_{21}^{*} & \wt g_{22}^{*}
\\ 
0 & \wt g_{21} & 0 & 0
\\
0 & \wt g_{22} & 0 & 0
\end{array}
\right)
\end{eqnarray}
and 
\begin{eqnarray}\label{LLL}
L=
\left(\begin{array}{cccc}
\sqrt{2\kappa} & 0 & 0 & 0
\\
0 & \sqrt{\gamma_{LC}} & 0 & 0
\\
0 & 0 & \sqrt{\gamma_{m,1}} & 0
\\
0 & 0 & 0 & \sqrt{\gamma_{m,2}}
\end{array}\right)\ .
\end{eqnarray}

The response of the system can be analyzed in terms of the output field operators defined by the relation 
\begin{eqnarray}
\va_{out}=\va_{in}-L\,\va\ .
\end{eqnarray} 
In the stationary regime we find $\va=-\pt{M+\epsilon\,V}^{-1}\ L\,\va_{in}$, so that 
the output operators can be expressed 
as 
\begin{eqnarray}\label{aoutSain}
\va_{out}=\SSS\,\va_{in}\ ,
\end{eqnarray}
where we have introduced the scattering matrix 
\begin{eqnarray}
\SSS=\id+L\,\pt{M+\epsilon\,V}^{-1} L
\end{eqnarray} 
which, at the first order in $\epsilon$, takes the form 
\begin{eqnarray}\label{SSS}
\SSS&\simeq&\SSS\oo-\epsilon\ L\,M^{-1}V\,M^{-1} L\ ,
\end{eqnarray} 
with
\begin{eqnarray}\label{SSSoo}
\SSS\oo&\equiv&\id+L\,M^{-1}L\ .
\end{eqnarray} 
Also these equations are easily adapted to the case with a single mechanical resonator by setting either $g_{j1}=0$ or $g_{j2}=0$ and considering $3\times 3$ matrices.

\subsection{The measurement}\label{measurement}

Here, we employ a phase sensitive detection strategy (see also App.\ref{App:detection})  which permits to  detect a quadrature
of the output field of the rf-resonator with the phase $\phi$, 
\begin{eqnarray}\label{X2out}
\hat{X}_{2}^{(out)}=\hat{a}_{2}^{(out)}\, e^{i\phi}+\hat{a}_2^{(out)\dagger}\,e^{-i\phi}\ ,
\end{eqnarray}
where the output annihilation operator is given by 
\begin{eqnarray}\label{a2outSaoin2}
\hat{a}_{2}^{(out)}=\pg{\SSS\,\va_{in}}_2 \ .
\end{eqnarray}
Including a finite detection efficiency $0<\eta<1$ (see App.~\ref{App:detection}),
the measured signal can be expressed as the time integration 
\begin{eqnarray}\label{hatm}
	\hat{m}=\int_{0}^{\tau} dt\,\pq{\sqrt{\eta}\,\,\hat{X}_{2}^{(out)}(t)+\sqrt{1-\eta}\,\hat{F}^{(noise)}(t)}
\end{eqnarray}
where $\tau$ is the detection time and $\hat{F}^{(noise)}$ is an operator describing additional detection noise (see App.~\ref{App:detection}). Here it is characterized by white noise correlations
$
\langle 
\hat{F}^{(noise)}(t)\ 
\hat{F}^{(noise)}(t')
\rangle
=\delta(t-t')$ 
and 
$\langle \hat{F}^{(noise)}\rangle=0$.
Note that the expression for the measurement operator $\hat m$~\rp{hatm} is obtained by a proper normalization of the integrated term, such that, when there is no signal (i.e. when the output mode described by $\hat{X}_{j}^{(out)}$ is in vacuum), the corresponding noise spectral density is equal to one (see App.~\ref{App:detection})~\cite{Zippilli}.

We want to probe the small perturbation $\epsilon\,V$, thus we are interested in the change of the detected signal $\hat m$ due to the perturbation
\begin{eqnarray}
\delta m=\av{\hat m}-\av{\hat m}_{\epsilon= 0}\ .
\end{eqnarray}
In the stationary regime, the value $\av{\hat{X}_{2}^{(out)}(t)}\equiv\av{\hat{X}_{2}^{(out)}}_{st}$ is time independent and we obtain (see Eq.~\rp{aoutSain})
\begin{eqnarray}
\av{\hat m}&=&\tau\ \sqrt{\eta}\ \av{\hat{X}_{2}^{(out)}}_{st}
\nn\\&=&
2\ \tau\ \sqrt{\eta}\ {\rm Re}\pq{ \beta\ e^{i\,\phi}\ \SSS_{2,2} }
\end{eqnarray}
so that 
\begin{eqnarray}
\delta m&=&2\ \tau\ \sqrt{\eta}\ {\rm Re}\pq{ \beta\ e^{i\,\phi}\ \pg{\SSS-\SSS\oo}_{2,2} }
\nn\\&\simeq&
-2\ \epsilon\ \gamma_{LC}\ \tau\ \sqrt{\eta}\ {\rm Re}\pq{ e^{i\,\phi}\ \beta
\pg{M^{-1}\ V\ M^{-1}}_{2,2} }\ ,
\end{eqnarray}
and by setting the homodyne phase to the value $\phi=\pi-\arg\pq{\beta
\pg{M^{-1}\ V\ M^{-1}}_{2,2}}$, we find
\begin{eqnarray}\label{deltam}
\delta m&\simeq&
2\ \epsilon\ \gamma_{LC}\ \tau\ \sqrt{\eta}\ \abs{\beta}
\abs{\pg{M^{-1}\ V\ M^{-1}}_{2,2} }\ ,
\end{eqnarray}
where an explicit expression for $M^{-1}$ can be obtained using the definition of $\SSS\oo$ in Eq.~\rp{SSSoo}, as
\begin{eqnarray}\label{Mm1}
M^{-1}=L^{-1}\pt{\SSS\oo-\id}L^{-1}\ .
\end{eqnarray}

Correspondingly, the 
variance of the measurement $
\Sigma^2=\av{\hat m^2}-\av{\hat m}^2$ can be expressed as
\begin{eqnarray}\label{Sigma}
\Sigma^2&=&\int_0^\tau\ \dd t_1\ \int_0^\tau\ \dd t_2\
\lpq{
\eta
\av{X_j\al{out}(t_1)\ X_j\al{out}(t_2)}_{\beta=0}
}\nn\\&&\rpq{
+
\pt{1-\eta}
\av{\hat F\al{noise}(t_1)\ \hat F\al{noise}(t_2)}
}
\nn\\&\simeq&
\tau\pq{\eta\ S_{X}\al{out,2}(0)+\pt{1-\eta}}\ ,
\end{eqnarray}
where we have introduced the power spectrum 
$S_{X}\al{out,2}(\omega)=
\int_{-\infty}^\infty\dd t\ e^{i\,\omega\,t}\av{\hat{X}_{2}^{(out)}(t)\ \hat{X}_{2}^{(out)}(0)}
$
of the rf output field quadrature, which is explicitly given by
\begin{eqnarray}\label{SXout2}
S_{X}\al{out,2}(0)\Bigr|_{\epsilon=0}
&=&
\sum_{j=1}^2
\pq{
\abs{\SSS_{2,j}}^2\pt{1+2\,\bar n_{aj}}
+ 
\abs{\SSS_{2,j+2}}^2\pt{1+2\,\bar n_{bj}}
}\ .\nn\\
\end{eqnarray} 
Specifically we find~\cite{Clerk} (see Eq.~\rp{SXout2}.)
\begin{eqnarray}\label{Sigma2}
\Sigma^2&=&\tau\,\eta
\sum_{j=1}^2
\pq{
\abs{\SSS_{2,j}}^2\pt{1+2\,\bar n_{aj}}
+ 
\abs{\SSS_{2,j+2}}^2\pt{1+2\,\bar n_{bj}}
}
\nn\\&&
+\tau\pt{1-\eta}\ .
\end{eqnarray}

\subsection{The signal-to-noise ratio}

In order to characterize the performance of the detection scheme, we analyze the signal-to-noise ratio 
$SNR=\delta m^{2}/\Sigma^{2}$. In particular we consider the lowest relevant order in $\epsilon$, which is given by
\begin{eqnarray}
SNR=
\frac{
\delta m^2
}{
\Sigma^2\Bigr|_{\epsilon=0}
}\ .
\end{eqnarray}

Relatively simple analytical expressions are found by assuming 
equal mechanical decay rates $\gamma_m=\gamma_{m,1}=\gamma_{m,2}$, 
and equal couplings, $g_{11}=g_{12}$ and $g_{21}=g_{22}$ (note that, since we employ a single rf driving field, when $\gamma_{m,1}=\gamma_{m,2}$, this last relation can be true only if the bare couplings are equal $g_{0,21}=g_{0,22}$).
We also introduce the optical and electrical cooperativities, which characterize the strength of the opto/electro-mechanical interaction relative to the rates of the dissipative processes,
\begin{eqnarray}\label{Gamma1}
\Gamma_1&=&\frac{2\,\abs{g_{1j}}^2}{\kappa\,\gamma_m}
\\
\Gamma_2&=&\frac{4\,\abs{g_{2j}}^2}{\gamma_{LC}\,\gamma_m}
\label{Gamma2}
\end{eqnarray}
for $j=1,2$, 
and we find the following expressions for the detection signal and noise [see Eqs.~\rp{deltam} and \rp{Sigma}]
\begin{eqnarray}\label{signal}
\delta m^2&=&16\,\tau^2\ \abs{\beta}^2\ \epsilon^2\ \eta\ \frac{
{\omega_{LC}^2}/{\gamma_{LC}^2}\ \pq{\pt{1+X}^2+Y^2}
}{
\pt{1+w}^4
}
\\
\rabs{\Sigma^2}_{\epsilon=0}&=&\tau\pq{
1+2\,\eta\ \frac{
\pt{1-w}^2\ \bar n_{a2}+u\ w
}{
\pt{1+w}^2
}
}\ ,
\label{noise}
\end{eqnarray}
where the parameters $X$ and $Y$ are small [they are related to the small parameters $\wt\omega_{LC}$, $\wt\Delta$ and  $\abs{\wt g_{2j}}$ introduced in Eqs.~\rp{gepsilon}-\rp{Deltaepsilon}] and we will show that they can be neglected (their explicit form is reported in App.~\ref{XY}). Moreover, in the 4-mode case,
\begin{eqnarray}\label{uu}
u&=& \frac{\bar n_{b1}+\bar n_{b2}}{\Gamma_1+\frac{1}{2}+\frac{2\,\delta^2}{\gamma_m^2}}
\pq{
1+\Gamma_1\pt{1-\cos\varphi}+\frac{\pt{
\Gamma_1\ \sin\varphi-\frac{2\,\delta}{\gamma_m}
}^2
}{
1+\Gamma_1\pt{1-\cos\varphi}
}
}
\nn\\
w&=&\Gamma_2\ \frac{
1+\Gamma_1\pt{1-\cos\varphi}
}{
\Gamma_1+\frac{1}{2}+\frac{2\,\delta^2}{\gamma_m^2}
}\ ,
\label{ww}
\end{eqnarray}
with $\varphi$ being related to the phases of the driving fields $\phi_j=\arg\pt{\EE_j}$ and $\phi_X=\arg\pt{V'}$ by 
\begin{eqnarray}\label{varphi}
\varphi=\phi_1-\phi_2-2\,\phi_X\ .
\end{eqnarray}
And, in the 3-mode case,
\begin{eqnarray}\label{u3}
u&=& \frac{4\ \bar n_{bj}}{\Gamma_1+1}
\\
w&=&\frac{\Gamma_2}{\Gamma_1+1}
\ .
\label{w3}
\end{eqnarray}
The corresponding signal-to-noise ratio takes the form 
\begin{eqnarray}
SNR&=&\label{SNRa}
\frac{
16\ \abs{\beta}^2\ \epsilon^2\ \tau\ \eta\
\omega_{LC}^2/\gamma_{LC}^2\ \pq{\pt{1+X}^2+Y^2}
}{
\pt{1+w}^2\pg{
\pt{1+w}^2+2\,\eta\pq{\pt{1-w}^2\ \bar n_{a2}+u\ w}
}
}\ .
\end{eqnarray}

\subsection{Optimization as a function of $w$}\label{Sec:maximization}

When $\omega_{LC}\gg\gamma_{LC}$ 
(that is the limit of our interest) the parameters $X$ and $Y$ [see Eqs.~\rp{X}, \rp{Y}, \rp{X3} and \rp{Y3}] are much smaller than one, so that they can be neglected, and,
as detailed in App.~\ref{maximization}, 
the maximum of $SNR$ is found when
\begin{eqnarray}\label{wopt}
w&=&\lpg{
\begin{aligned}
&\frac{3}{2}\pq{\frac{1}{3}-\frac{\varrho}{\sigma}+\sqrt{\pt{1-\frac{\varrho}{\sigma}}\pt{\frac{1}{9}-\frac{\varrho}{\sigma}}}
}  & \text{for}\  \frac{\varrho}{\sigma}<\ovl\xi\ ,\\
&0  & \text{for}\  \frac{\varrho}{\sigma}\geq\ovl\xi\ ,
\end{aligned}
}\nn\\
\end{eqnarray}
where
\begin{eqnarray}\label{varrho}
\varrho&=&{1+\eta\ u/2}\ ,
\\
\sigma&=&{1+2\,\eta\ \bar n_{a2}}\ ,
\label{sigma}
\end{eqnarray}
and (see App.~\ref{maximization})
\begin{eqnarray}
\ovl\xi\sim 0.097
\end{eqnarray}
is the value which separates the regimes in which either our system or a simple LC resonator is more efficient for sensing. Specifically, the condition $w=0$ (which maximizes $SNR$ when $\frac{\varrho}{\sigma}>\ovl\xi$) is achieved when the electromechanical cooperativity is $\Gamma_2=0$ or, in other terms, when there is no interaction between the LC and the mechanical resonators.
The corresponding maximum of $SNR$ is (for $\omega_{LC}\gg\gamma_{Lc}$)
\begin{eqnarray}\label{SNRmax}
SNR_{\rm max}&= &
\lpg{
\begin{aligned}
&\frac{
12\ \abs{\beta}^2\ \epsilon^2\ \tau\ \eta\
\omega_{LC}^2/\gamma_{LC}^2
}{
\sigma
\pt{1-\frac{\varrho}{\sigma}}^2\pg{
1-\pq{\frac{9}{4}\ 
\pt{\frac{1}{9}-\frac{\varrho}{\sigma}}\
\pt{1+\sqrt{
\frac{1-\frac{\varrho}{\sigma}}{\frac{1}{9}-\frac{\varrho}{\sigma}}
}}
}^2
}
} 
\\
& & & \hspace{-2.3cm}\text{for} \  \frac{\varrho}{\sigma}<\ovl\xi \ ,
\\
& SNR_0
& & \hspace{-2.3cm}\text{for}\  \frac{\varrho}{\sigma}\geq\ovl\xi \ .
\end{aligned}
}\nn\\
\end{eqnarray}
where 
\begin{eqnarray}\label{SNR0}
SNR_0&\equiv& SNR\Bigl|_{w=0}
\nn\\
&=&
16\ \abs{\beta}^2\ \epsilon^2\ \tau\ \eta
\ \frac{\omega_{LC}^2}{\gamma_{LC}^2}\
\frac{1}{\sigma}
\end{eqnarray}
is the signal-to-noise ratio obtained with a simple LC circuit.
For $\varrho/\sigma<\ovl\xi$, the value of $SNR_{\rm max}$ 
decreases monotonically with increasing $\varrho$, that is with increasing $u$ [see Eq.~\rp{varrho}] or, in other terms, it decreases with the number of thermal mechanical excitations $\bar n_{bj}$ and increases with the optomechanical cooperativity $\Gamma_1$ [see Eqs.~\rp{uu} and \rp{u3}]. In turn, $SNR_{\rm max}$ 
decreases monotonically also with $\sigma$, i.e. with the number of thermal rf excitations $\bar n_{a2}$ [see Eq.~\rp{sigma}]. This indicates that better signal-to-noise ratio is achieved for both low mechanical and rf noise.

\subsection{The relative signal-to-noise ratio}\label{relSNR}

However, what interests us here is whether and when we can employ our opto-electromechanical system to achieve enhanced sensing with respect to a simple LC resonator, that corresponds to the condition $w=0$ (i.e. $\Gamma_2=0$) and for which the signal-to-noise ratio is given by Eq.~\rp{SNR0}. Therefore, we characterize the efficiency of our system in terms of the relative signal-to-noise ratio 
\begin{eqnarray}\label{r}
r&\equiv&\frac{SNR}{SNR_0}
\nn\\
&=&
\frac{
1+2\,\eta\ \bar n_{a2}
}{
\pt{1+w}^2\pg{
\pt{1+w}^2+2\,\eta\pq{\pt{1-w}^2\ \bar n_{a2}+u\ w}
}
}\ ,
\end{eqnarray}
where we neglected the parameters $X$ and $Y$ in the definition of $SNR$~\rp{SNRa}.
In particular, the value of $r$ is maximized when $w$ is given by Eq.~\rp{wopt}, with maximum value equal to 
\begin{eqnarray}\label{rmax}
r_{\rm max}&= &
\lpg{
\begin{aligned}
&\frac{
3
}{
4\ 
\pt{1-\frac{\varrho}{\sigma}}^2\pg{
1-\pq{\frac{9}{4}\ 
\pt{\frac{1}{9}-\frac{\varrho}{\sigma}}\
\pt{1+\sqrt{
\frac{1-\frac{\varrho}{\sigma}}{\frac{1}{9}-\frac{\varrho}{\sigma}}
}}
}^2
}
} 
\\
& & & \hspace{-1.5cm}\text{for} \  \frac{\varrho}{\sigma}<\ovl\xi \ ,
\\
& 1
 & & \hspace{-1.5cm}\text{for}\  \frac{\varrho}{\sigma}\geq\ovl\xi \ .
\end{aligned}
}\nn\\
\end{eqnarray}
Enhanced sensing corresponds to $r>1$, that can be obtained when $\frac{\varrho}{\sigma}<\ovl\xi$. In particular,  
the maximum~\rp{rmax} decreases monotonically with increasing $\frac{\varrho}{\sigma}$ (see Fig.~\ref{fig0.0}), indicating that the enhancement is more pronounced for small $\varrho$ (i.e. small mechanical noise $\bar n_{bj}$ and large optomechanical cooperativity $\Gamma_1$) and large $\sigma$ (i.e. high rf noise $\bar n_{a2}$).

\subsection{Optimization of the 4-modes model as a function of $\delta$ and $\varphi$}\label{varphidleta4modes}

We have seen that, in general, larger signal-to-noise ratio is obtained for smaller values of $u$ which accounts for the mechanical thermal noise [see Eqs.~\rp{uu} and \rp{u3}].
In the case of the 4-modes model, $u$ is minimized, as a function of $\delta$ for 
\begin{eqnarray}\label{deltaopt}
\delta_{opt}=\gamma_m\ \pt{\frac{1}{2}+\Gamma_1}\ \frac{\sin\varphi}{1+\cos\varphi}
\end{eqnarray}
such that the corresponding minimum value is
\begin{eqnarray}\label{u4}
u(\delta_{opt})&=& \frac{
\bar n_{b1}+\bar n_{b2}
}{
\Gamma_1+\frac{1}{2}
}\ .
\end{eqnarray}
Correspondingly, $w$ takes the form
\begin{eqnarray}\label{w4}
w(\delta_{opt})=\frac{\Gamma_2}{\Gamma_1+\frac{1}{2}}\ 
\frac{1+\cos\varphi}{2}\ .
\end{eqnarray} 
Enhanced sensing can be found for a finite value of $w$, see Eq.~\rp{wopt}. 
This implies that the smallest possible value of the electromechanical cooperativity $\Gamma_2$, which maximizes the $SNR$, occurs when $\varphi=0$ (that, according to the optimal condition~\rp{deltaopt}, corresponds to $\delta=0$).

\subsection{Impedance matching regime}\label{impedance-matching}

Enhanced sensing can be achieved for values of $w$ not far from $w\sim 1$ [see Eq.~\rp{wopt}]. In fact, the optimal value of $w$ in Eq.~\rp{wopt}, decreases monotonically with increasing $\frac{\varrho}{\sigma}$ (for $0<\frac{\varrho}{\sigma}<\ovl\xi$), and it takes the values $w=1$ when $\frac{\varrho}{\sigma}=0$ and $w\sim 0.52$ when $\frac{\varrho}{\sigma}=\ovl\xi$. In particular, the largest enhancement [in terms of the relative signal-to-noise ratio~\rp{r}] can be achieved  in the limit $\frac{\varrho}{\sigma}\to 0$, that corresponds to $w\to 1$.

We highlight that the condition $w=1$ has a specific physical meaning.
It corresponds to setting to zero the contribution of the electromagnetic noise (the term proportional to $\bar n_{a2}$) in the expression for the noise~\rp{noise}. This is the regime of perfect impedance matching where the reflection of input rf-fields is completely suppressed (in absence of the perturbation, $\epsilon=0$). It corresponds to the parameter regime for which the element $(2,2)$ of the scattering matrix~\rp{SSSoo}, which corresponds to the reflection of rf-fields, is zero, $\pg{\SSS\oo}_{2,2}=0$ [see Eq.~\rp{SXout2} and \rp{Sigma2}]. In fact, in general 
\begin{eqnarray}\label{SImpMatch}
\pg{\SSS\oo}_{2,2}=\frac{w-1}{w+1}\ .
\end{eqnarray}

We also note that 
the output rf noise~\rp{SXout2} (for $\epsilon=0$), takes the general form
\begin{eqnarray}
S_X\al{out,2}=\pt{\frac{w-1}{w+1}}^2\pt{1+2\,\bar n_{a2}}
+\frac{2\ w\pt{2+u}}{\pt{w+1}^2}
\end{eqnarray}
So that when $w=1$ it reduces to
\begin{eqnarray}
S_X\al{out,2}\Bigl|_{w=1}=1+\frac{u}{2}
\end{eqnarray}
which exceeds the vacuum noise level (that is equal to 1) by the parameter $u$ [defined in Eqs.~\rp{uu}, \rp{u3} and \rp{u4}] that expresses the contribution of the mechanical noise reduced by the effect of laser cooling. 
This means that in the limit of large optomechanical cooperativity $\Gamma_1$
the output rf noise approaches the vacuum noise level.
In other terms, in this regime the rf noise is rerouted to the unused optical port and the mechanical noise is strongly reduced by the effect of laser cooling. What remains in the rf output is the contribution of the optical noise which, however, is essentially vacuum noise at any realistic temperature.  
This is the main mechanism at the basis of the enhanced sensing in this system.

\begin{figure}[t]
\centering
\includegraphics[width=0.48\textwidth]{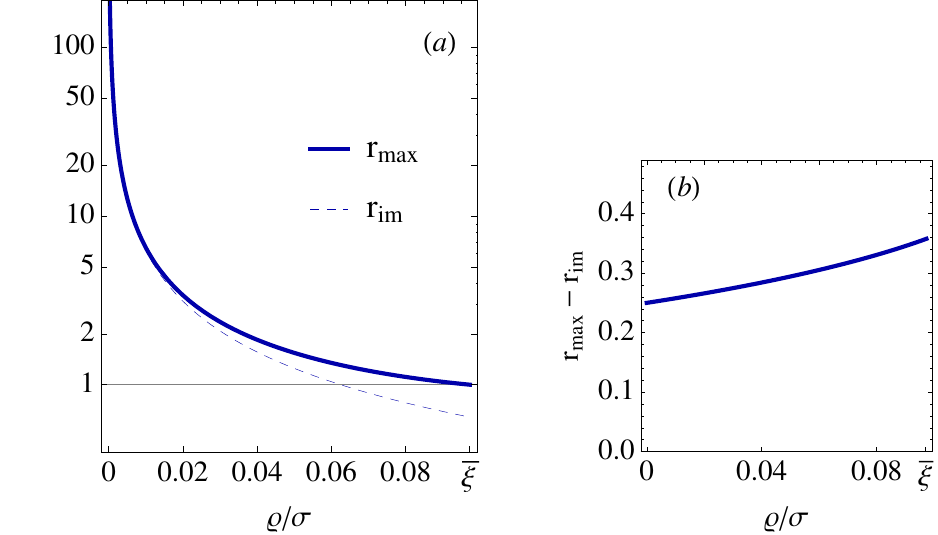}
\caption{
(a) Optimized relative signal-to-noise $r_{\rm max}$, Eq.~\rp{rmax}, and corresponding result in the impedance matching regime ($w=1$) $r_{\rm im}$, Eq.~\rp{rim}, as a function of the $\varrho/\sigma$ [see Eqs.~\rp{varrho} and \rp{sigma}]. (b) corresponding difference $r_{\rm max}-r_{\rm im}$.
The results are plotted for $0<\varrho/\sigma<\ovl\xi$, that is the complete range over which we can observe enhanced sensing ($r>1$).
$r_{\rm max}$ diverges as $\frac{1}{\varrho/\sigma}$. 
$r_{\rm im}$ is a good approximation of $r_{\rm max}$ at small values of $\varrho/\sigma$.
}
\label{fig0.0}
\end{figure}

To summarize, the relative signal-to-noise ratio (see Sec.~\ref{relSNR}) 
can be optimized when $\varrho/\sigma$ [see Eqs.~\rp{varrho} and \rp{sigma}] is small, by tuning the cooperativities to the values for which $w\sim 1$ [see Eqs.~\rp{ww} and \rp{w3}].
According to Eq.~\rp{SImpMatch}, this corresponds to the regime in which the system is operated close to the conditions of impedance matching, where the reflection of rf-fields is suppressed. 
When $w=1$ the signal-to-noise ratio reduces to the form  
\begin{eqnarray}\label{SNRb}
SNR_{\rm im}&\equiv&SNR\Bigl|_{w=1}
\nn\\&=&
\abs{\beta}^2\ \epsilon^2\ \tau\ \eta\ 
\frac{\omega_{LC}^2}{\gamma_{LC}^2}\ 
\frac{1}{\varrho}\ ,
\end{eqnarray} 
and the corresponding expression for the relative signal-to-noise ratio is
\begin{eqnarray}\label{rim}
r_{\rm im}&\equiv&\frac{SNR_{\rm im}}{SNR_0}
\nn\\
&=&\frac{1}{16}\ 
\frac{\sigma}{\varrho}\ ,
\end{eqnarray}
with $SNR_0$, $\varrho$ and $\sigma$ defined, respectively, in Eqs.~\rp{SNR0}, \rp{varrho} and \rp{sigma}.
These equations are good approximations of the general results~\rp{SNRa} and \rp{r}when $\varrho/\sigma$ [see Eqs.~\rp{varrho} and \rp{sigma}] is small (see Fig.~\ref{fig0.0}).

\subsubsection{The 4-modes model in the nonreciprocal regime}\label{nonreciprocal}

As discussed in Ref.~\cite{Eshaqi}, this system can be operated in a nonreciprocal regime where the transmission from rf-to-optical port is suppressed, whereas the reversed transmission from optical-to-rf is large.
Nonreciprocity has been discussed as a tool to enhance sensing~\cite{Clerk,mcdonald2020,Nori}, it is therefore interesting to verify whether in our case, nonreciprocity can be exploited to maximize the signal-to-noise ratio.

Ref.~\cite{Eshaqi} demonstrated that
the transmission from the rf input to the optical output is completely suppressed ($\pg{\SSS\oo}_{1,2}=0$) when the phase $\varphi$~\rp{varphi} fulfills the relation
$\ee^{\ii\ \varphi}=-\frac{\gamma_{m}-2\,i\,\delta}{\gamma_{m}+2\,i\,\delta}$. At the same time the transmission from the optical input to the rf output ($\pg{\SSS\oo}_{2,1}$) is maximized when the cooperativities are equal, $\Gamma_1=\Gamma_2\equiv\Gamma$
and $\delta=\frac{\gamma_m}{2}\sqrt{2\, \Gamma-1}$. Under these conditions it is easy to verify that $w=1$ [see Eq.~\rp{ww}], indicating that this is a particular case
of the impedance matching regime,
and also in this case the signal-to-noise ratio is equal to Eq.~\rp{SNRb}. As we have seen, this is not the optimal regime for sensing. But it approaches the optimal result when $\varrho/\sigma$ [see Eqs.~\rp{varrho} and \rp{sigma}] is small (see also Sec.~\ref{varphidleta4modes}).

\subsection{
3-modes vs 4-modes models}

The two configurations are essentially equivalent, in the sense that one can be mapped to the other by appropriately rescaling the system parameters such that the values of $u$ and $w$ are equal in the two scenarios. 
However it is reasonable to ask whether one is more efficient than the other at give mechanical frequencies. Specifically, we assume similar large cooperativities such that, on the one hand, the values of $w$ (Eq.~\rp{w3} and Eq.~\rp{w4} with $\varphi=0$) are similar in the two cases, and on the other, according to Eqs.~\rp{u3} and \rp{u4}, only the mechanical thermal occupancies, which are determined by the mechanical frequencies, are relevant in this comparison.

We introduce the symbols
$u_\xi$ with $\xi=3,4$ to distinguish the parameters corresponding to the two cases. Namely, we define $u_3$ to be equal to Eq.~\rp{u3}, while $u_4$ to Eq.~\rp{u4}.
The 4-mode case is more efficient than the 3-mode one when $u_4<u_3$ and vice-versa.
Now, in the 4-modes scenario, one mechanical resonator exhibits a frequency higher and the other lower than the LC resonance frequency, $\omega_{LC}$ (in fact, from~\rp{resonance}, we find $\omega_{LC}=\pt{\omega_{m1}+\omega_{m2}}/{2}$). 
Hereafter we assume that $\omega_{m1}$ is the smallest one
\begin{eqnarray}
\omega_{m1}<\omega_{LC}=\frac{\omega_{m1}+\omega_{m2}}{2}<\omega_{m2}\ .
\end{eqnarray}
Correspondingly we find
\begin{eqnarray}
2\bar n_{b1}>\bar n_{b1}+\bar n_{b2}>2\bar n_{b2}\ .
\end{eqnarray}
This leads us to consider two cases in our analysis of the 3-modes model: one where the LC frequency exceeds the mechanical frequency, and the other where it is lower [see Figs.~\ref{fig1.2} (a) and (b)].
Thus, the 3-modes model with the low frequency mechanical resonator is always less efficient than the 4-modes model (i.e. $u_3>u_4$). 
On the contrary the 3-modes model with the high frequency mechanical resonator can result more efficient (i.e. it is possible to have $u_3<u_4$). Specifically, this is observed when 
$4\bar n_{b2}<\bar n_{b1}+\bar n_{b2}$, that is (at any realistic temperature at MHz frequencies)
when $\omega_{m2}>3\omega_{m1}$ or, in other terms the 3-modes model is more efficient than the 4-modes one when
\begin{eqnarray}\label{omegaLCm2}
\omega_{LC}<\frac{2}{3}\ \omega_{m2}\ .
\end{eqnarray}
This shows that, although the two cases are essentially equivalent, one may prefer one over the other depending on the specific values of the mode frequencies.

\section{
Results}\label{results}

\begin{figure}[t]
\centering
\includegraphics[width=0.48\textwidth]{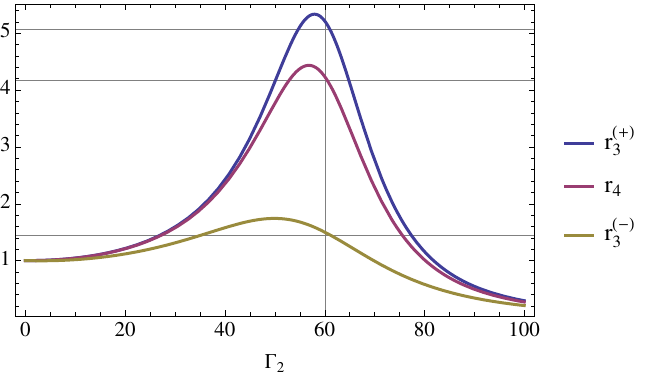}
\caption{
Relative signal-to-noise ratio [Eq.~\rp{r}] as a function of the electromechanical cooperativity $\Gamma_2$, when the optomechanical  cooperativity is fixed at the value indicated by the vertical thin line $\Gamma_1=60$.
The central curve ($r_4$) corresponds to the 4-modes model. The upper ($r_3\al{+}$) and lower ($r_3\al{-}$) curves correspond to the 3-modes models achieved by considering, respectively, only the upper and lower frequency mechanical mode.
The horizontal thin solid lines indicate the values achieved under the conditions of perfect impedance matching according to Eq.~\rp{SNRb}. 
The other system parameters are $\omega_{LC}=5$MHz, $\gamma_{LC}=6$kHz, $\omega_1=2$MHz, $\omega_2=8$MHz, $\gamma_m=500$Hz, $\delta=0$, $\varphi=0$. The temperature is $T=0.1$K, corresponding to $\bar n_{a2}=2618$, $\bar n_{b1}=6545$, $\bar n_{b2}=1636$. $\eta=0.09$. 
}
\label{fig0.1}
\end{figure}

\begin{figure}[t]
%\flushleft
\centering
\includegraphics[width=0.48\textwidth]{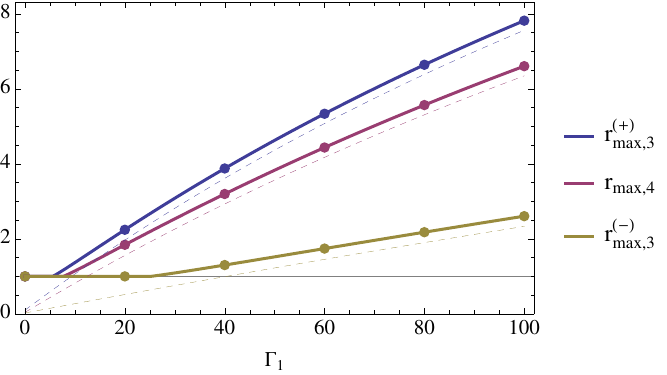}
\caption{
Optimized relative signal-to-noise ratio [Eqs.~\rp{rmax} and \rp{rim}] as a function of the optical cooperativity $\Gamma_1$. The solid lines correspond to Eq.~\rp{rmax} so that they are evaluated for the values of the electromechanical cooperativity, $\Gamma_2$, which make the parameter $w$ equal to Eq.~\rp{wopt} for each value of $\Gamma_1$. The dashed lines correspond to the condition of perfect impedance matching [Eq.~\rp{rim}]: they are evaluated for the values of $\Gamma_2$ which make the parameter $w=1$ for each value of $\Gamma_1$. 
The dots are evaluated by a numerical maximization of $SNR/SNR_0$, as a function of $\Gamma_2$ for each value of $\Gamma_1$, using the full expression of $SNR$~\rp{SNRa}, which includes the parameters $X$ and $Y$ evaluated assuming equal bare opto/electro-mechanical couplings (note that the frequency shift $\wt\Delta$~\rp{Deltat}, contained in the definition of $Y$~\rp{Y}, depends on the bare couplings and can not be expressed simply in terms of the cooperativities). The coincidence between dots and solid lines verifies the validity of neglecting the parameters $X$ and $Y$ in Eq.~\rp{SNRa} and it is a further justification of our analytical maximization (see App.~\ref{maximization}) based only on the minimization of the denominator of $SNR$ (see also Figs.~\ref{fig4} and \ref{fig5}). 
Upper, central and lower curves correspond, respectively, to the 3-modes model with the upper frequency mechanical mode, the 4-modes model and the 3-modes model with the lower frequency mechanical mode.
The thin black horizontal line at the value of 1 is intended to emphasize the values for which our protocol is efficient ($r_{\rm max}>1$). 
The other parameters are as in Fig.~\ref{fig0.1}.
}
\label{fig0.2}
\end{figure}

\begin{figure}[t!]
\centering
\includegraphics[width=0.45\textwidth]{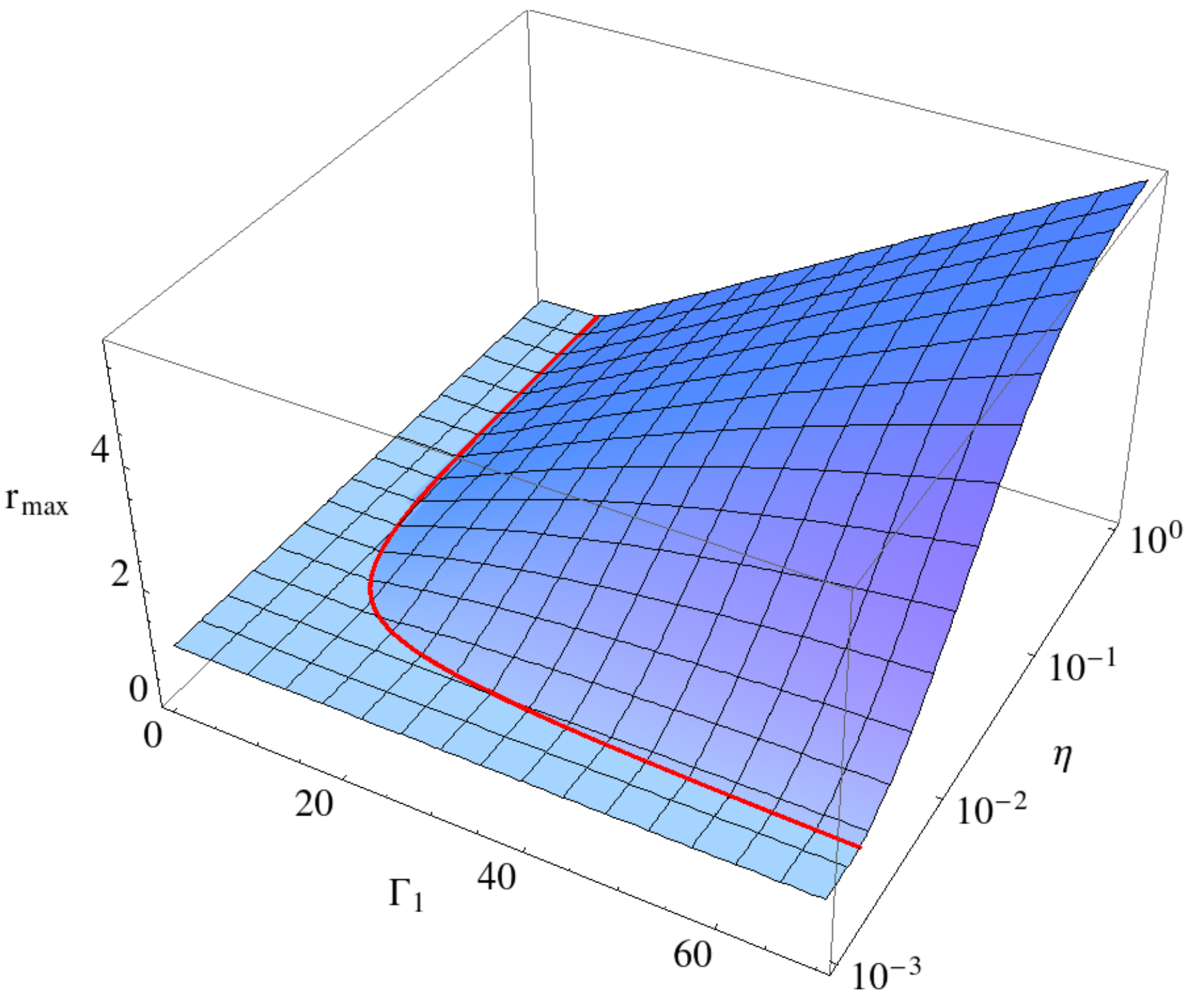}
\caption{
Optimized relative signal-to-noise ratio [Eq.~\rp{rmax}] for the 4-modes model as a function of the optomechanical cooperativity parameter $\Gamma_1$ 
[see Eq.~\rp{Gamma1}] and the detection efficiency $\eta$ [introduced in Eq.~\rp{hatm}].
The red line separates the region where $r_{\rm max}>1$ from that  where
$r_{\rm max}=1$. The system parameters are as in Fig.~\ref{fig0.1}.
}
\label{fig2}
\end{figure}

\begin{figure}[t!]
\centering
\includegraphics[width=0.45\textwidth]{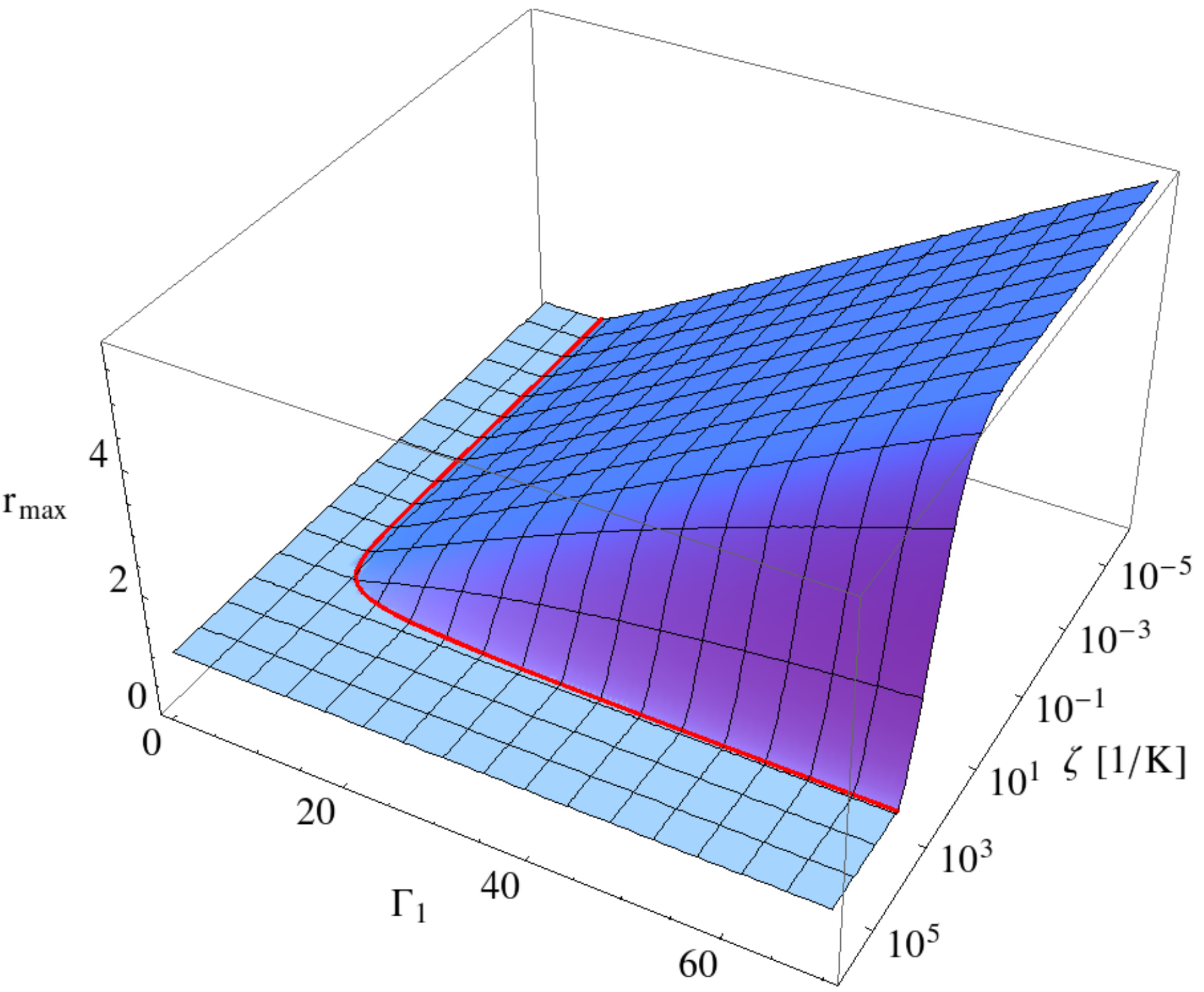}
\caption{
Optimized relative signal-to-noise ratio [Eq.~\rp{rmax}] for the 4-modes model as a function of the optomechanical cooperativity parameter $\Gamma_1$ [see Eq.~\rp{Gamma1}] and the detection noise parameter $\zeta$ which determines the detection efficiency according to Eq.~\rp{etaZeta}.
The other parameters and line styles are as in Fig.~\ref{fig2}.}
\label{fig3}
\end{figure}

\begin{figure}[t]
\flushleft
\includegraphics[width=0.45\textwidth]{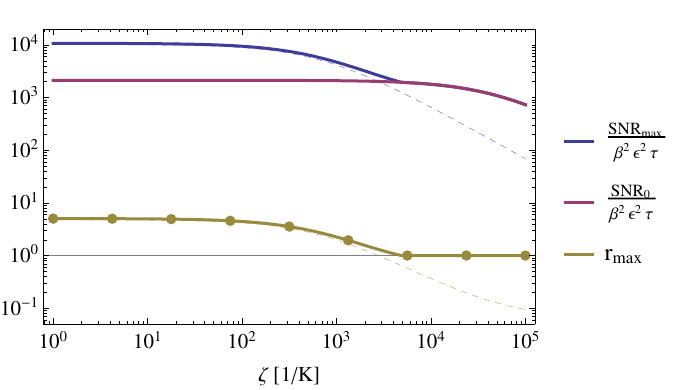}
\caption{Analysis of the signal-to-noise ratio as a function of the detection noise parameter $\zeta$, for the 4-modes model with $\Gamma_1=60$.
The three solid thick curves correspond to the quantities reported in the inset. 
The dashed thin lines are the corresponding results in the impedance matching regime, Eqs.~\rp{SNRb} and \rp{rim}. The dots are evaluated by a numerical maximization of $SNR/SNR_0$, using the full expression of $SNR$~\rp{SNRa}, as a function of $\Gamma_2$ for each value of $\zeta$.
The thin black horizontal line at the value of 1 is intended to emphasize the values for which our protocol is efficient, i.e. for which the lower thick line is larger than one ($r_{\rm max}>1$). 
The other parameters are as in Fig.~\ref{fig0.1}.
}
\label{fig4}
\end{figure}

\begin{figure}[t]
\flushleft
\includegraphics[width=0.45\textwidth]{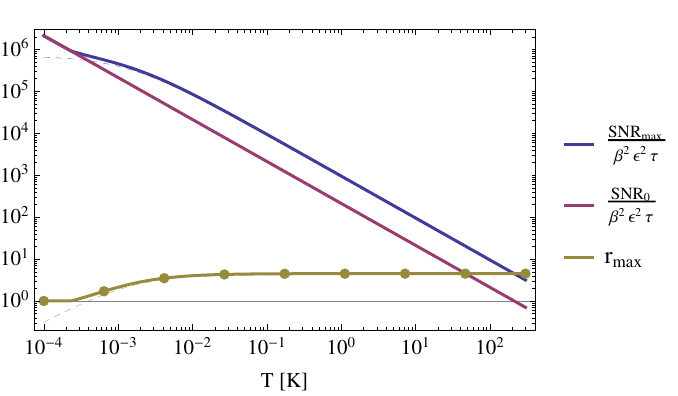}
\caption{
Analysis of the signal-to-noise ratio as a function of the temperature $T$, for the 4-modes model with $\zeta=100\,$K$^{-1}$ and 
$\Gamma_1=60$.
The other parameters and the line styles are as in Fig.~\ref{fig4}.
}
\label{fig5}
\end{figure}

\begin{figure}[t]
\flushleft
\includegraphics[width=0.48\textwidth]{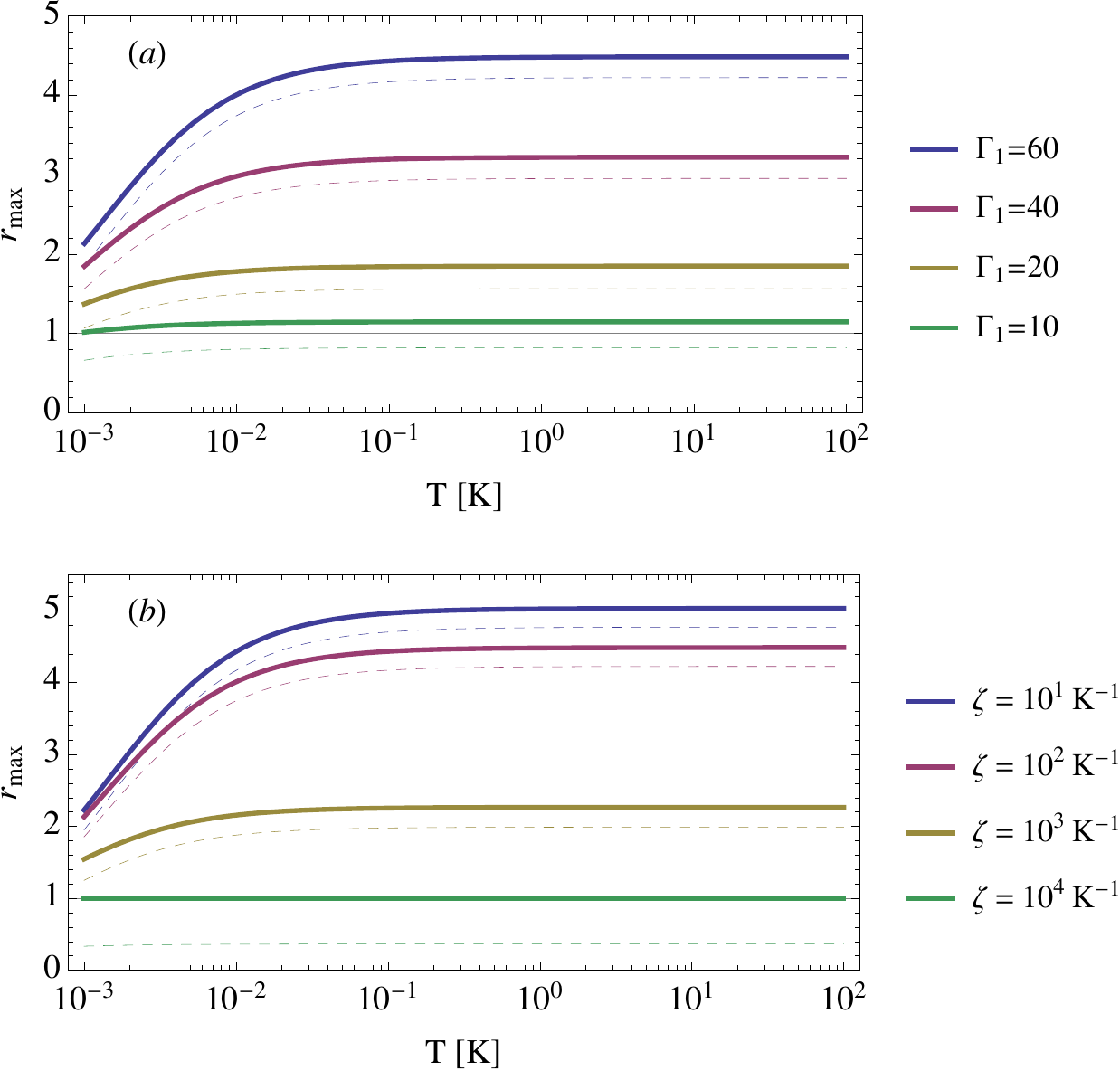}
\caption{
Optimized reduced signal-to-noise ratio for the 4-modes model as a function of the temperature $T$ and for various values of (a) the optomechanical cooperativity $\Gamma_1$ and (b) the noise parameter $\zeta$ (see Eq.~\rp{etaZeta}). In (a) $\zeta=100\,$K$^{-1}$ and in (b) $\Gamma_1=60$.
The thick solid lines are the results for $r_{\rm max}$, Eq.~\rp{rmax}, and the thin dashed lines for  $r_{\rm im}$, Eq.~\rp{rim}.
The other parameters are as in Fig.~\ref{fig4}.
}
\label{fig6}
\end{figure}

\begin{figure}[t]
\centering
\includegraphics[width=0.45\textwidth]{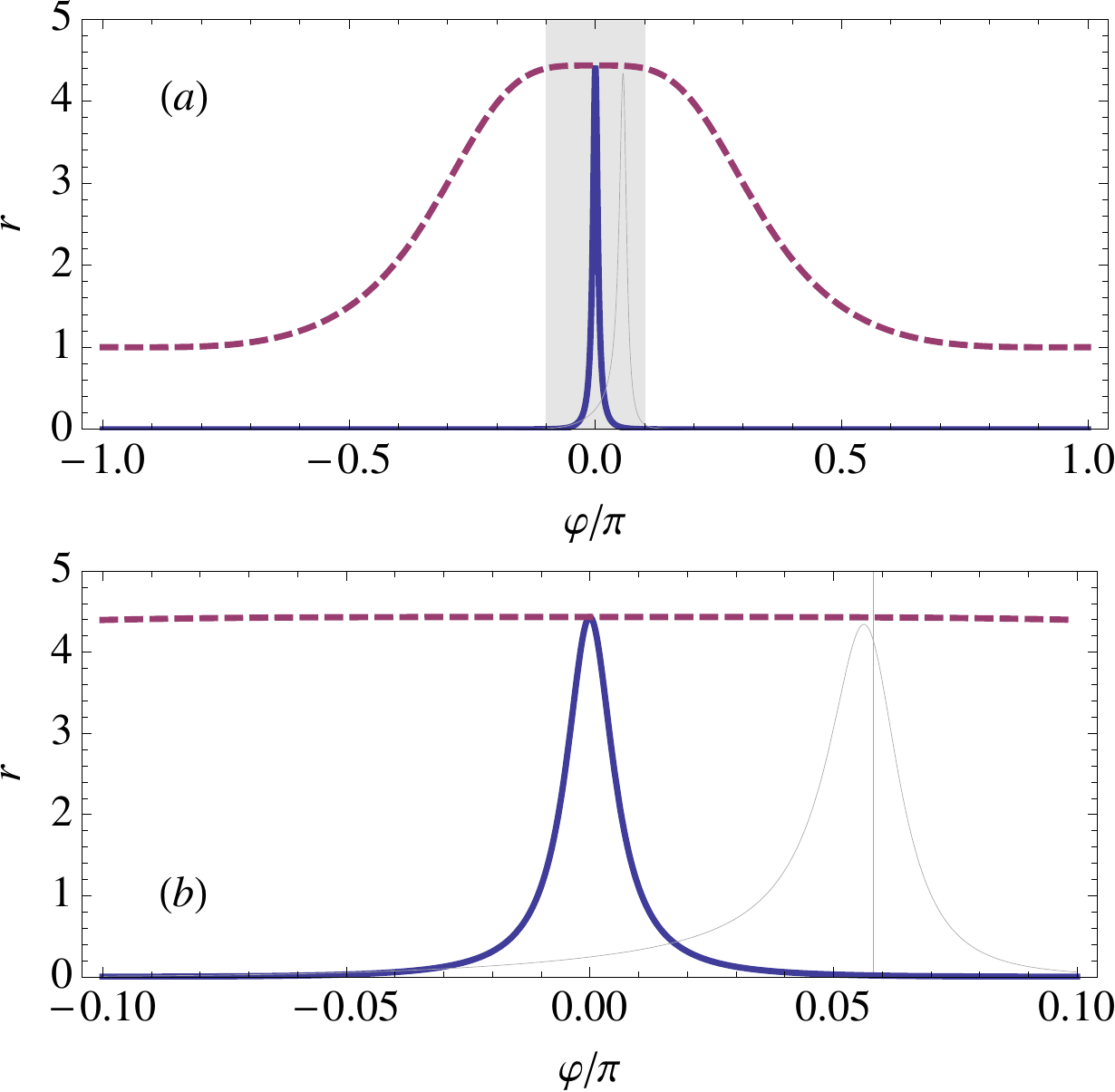}
\caption{
(a) Relative signal-to-noise ratio $r$, Eq.~\rp{r},  for the 4-modes model as a function of the phase $\varphi$, Eq.~\rp{varphi}. Panel (b) is a zoom in on the gray area of (a).
The thick solid blue lines and the thick dashed red lines are obtained by setting the electromechanical cooperativity $\Gamma_2$ to the value which maximizes the signal-to-noise ratio for $\varphi=0$ and $\delta=0$, according to Eqs.~\rp{wopt} and \rp{ww}. The thick solid blue lines are then evaluated at fixed $\delta=0$. The thick dashed red lines are evaluated by varying the value of $\delta$ according to Eq.~\rp{deltaopt}.
The thin gray lines are evaluated for the values of the cooperativities and of $\delta$ which correspond to the nonreciprocal regime described in Sec.~
\ref{nonreciprocal}~\cite{Eshaqi}. In particular, the vertical thin line in (b) indicates the value of $\varphi$ at which the system is nonreciprocal.
The other parameters are as in Fig.~\ref{fig0.1}.
}
\label{fig0.3}
\end{figure}

Let us now focus on specific physical parameters. We assume cryogenic temperatures, high quality factors resonators (see caption of Fig.~\ref{fig0.1}) and other parameters consistent with those used in Ref.~\cite{Eshaqi}, which deals with a similar setup (see also Sec.~\ref{nonreciprocal}), for which we have already verified the validity of all the approximations at the basis of our models.

We consider mechanical and rf frequencies of few MHz for which, in particular, Eq.~\rp{omegaLCm2} is true. In this case the 3-modes model is more efficient than the 4-modes one, when the former employs the higher mechanical frequency of the latter. This is shown in Figs.~\ref{fig0.1} and \ref{fig0.2}. These plots show comparisons of the relative signal-to-noise ratio $r$ (in Fig.~\ref{fig0.1}) and of the optimized values $r_{\rm max}$ (in Fig.~\ref{fig0.2}) for both the 4-modes and 3-modes models, distinguishing the corresponding results by the subscripts $4$ and $3$. In particular, in the case of the 3-modes models we further distinguish, with the superscript $(+)$ and $(-)$, the results achieved by considering either the high or low frequency mechanical resonator. 
In Fig.~\ref{fig0.1} the maxima as a function of the electromechanical cooperativity $\Gamma_2$ correspond to the values which make the corresponding parameter $w$ [see Eq.~\rp{ww} and \rp{w3}] equal to the optimal value~\rp{wopt}, and which is always slightly smaller than the optomechanical cooperativity. 
The result in Fig.~\ref{fig0.2} are evaluated as a function of the optomechanical cooperativity $\Gamma_1$ by varying also the electromechanical cooperativity $\Gamma_2$ in such a way that the optimal condition~\rp{wopt} is always fulfilled. This figure reports also a comparison of the optimized results $r_{\rm max}$~\rp{rmax} (solid lines) with the corresponding results obtained in the impedance matching regime $r_{\rm im}$~\rp{rim} (dashed lines), showing that the two results are always very close to each other. 

The dots in Fig.~\ref{fig0.2} (see also Figs.~\ref{fig4} and \ref{fig5}) are evaluated by the numerical maximization of the full expression of $SNR$~\rp{SNRa} which include also the parameters $X$ and $Y$, see App.~\ref{XY}, that we neglected in the other curves. These results show that neglecting these parameters is a valid approximation.

\subsection{Effect of the detection efficiency}

In the following, we investigate, in more detail, the effect of the detection noise and of the temperature on the sensing efficiency in our system. Specifically, the results presented hereafter are evaluated for the 4-modes model. However it is clear that similar results (apart from the results as a function of the $\varphi$~\rp{varphi} in Fig.~\ref{fig0.3}, which are specific to the 4-modes model) are valid also for the 3-modes model.

We observe that in Sec.\ref{relSNR} we have shown that our protocol is efficient ($r_{\rm max}<1$), if $\varrho/\sigma<\ovl\xi$. This condition can be equivalently expressed as a condition over the detection efficiency $\eta$, such that we can state that our sensing protocol can be efficient  if 
\begin{eqnarray}\label{etacond}
\eta>\frac{1-\ovl\xi}{2\ \bar n_{a2}-u/2}\ .
\end{eqnarray}
The optimized relative signal-to-noise ratio $r_{\rm max}$~\rp{rmax} is reported in Fig.~\ref{fig2} as a function of the cooperativity $\Gamma_1$ [see Eq.~\rp{Gamma1}] and the detection efficiency $\eta$ introduced in Eq.~\rp{hatm} (see also App.\ref{App:detection}). This figure shows that an efficient detection ($r>1$) is obtained for sufficiently large cooperativity 
(which minimizes $u$) and sufficiently large detection efficiency in agreement with the constraint of Eq.~\rp{etacond}. Moreover, we observe that a lower detection efficiency $\eta$ requires a larger cooperativity for achieving an improved detection.

In order to analyze the effectiveness of the detection strategy proposed here as a function of the temperature and of the detection noise, it is important to highlight that, according to our definitions, the detection efficiency parameter $\eta$ itself introduced here depends upon the temperature (see App.\ref{App:detection}). In phase sensitive detection strategies, the signal field is mixed with a reference field. The resulting field is then measured and information about the signal quadratures are extracted. The inevitable noise added to the signal by the detection process can be characterized by the added noise power relative to the coherent component of the reference field, which under general conditions, can be considered proportional to the temperature, and we express it as $\zeta\,T$. As discussed in the appendix~\ref{App:detection}, the corresponding detection efficiency can be expressed as
\begin{eqnarray}\label{etaZeta}
\eta=\frac{1}{1+\zeta\ T}\ .
\end{eqnarray}
Now, since for rf frequencies $\bar n_{a2}\sim\frac{k_B\ T}{\hbar\,\omega_{LC}}$ and $\bar n_{bj}\sim\frac{k_B\ T}{\hbar\,\omega_j}$, we obtain
that $u$ [see Eqs.~\rp{uu} and \rp{u3}] is proportional to the temperature, and we can write 
\begin{eqnarray}
u\sim\check u\ \frac{k_B\ T}{\hbar}
\end{eqnarray}
with
\begin{align}
\check u&=\frac{1}{\Gamma_1+1}\ \frac{4}{\omega_j} & &\text{for the 3-modes model}\ ,
\\
\check u&=\frac{1}{\Gamma_1+\frac{1}{2}}\ \frac{2\ \omega_{LC}}{\omega_1\ \omega_2} & &\text{for the 4-modes model, with }\ \delta=\varphi=0\ ,
\end{align}
where, in agreement with Eq.~\rp{resonance}, we have used the relation $\omega_{LC}=(\omega_1+\omega_2)/2$.
As a result we find that our protocol is efficient (i.e. $r>1$) when $\zeta$ is sufficiently small, namely when 
\begin{eqnarray}
\label{Zetan}
\zeta<\frac{2\ k_B}{
\hbar\,\omega_{LC}}\frac{\ovl\xi}{1-\ovl\xi}
\pt{1-\frac{\omega_{LC}\ \check u}{ 4\ \ovl\xi}}-\frac{1}{T}
\ .
\end{eqnarray}
These results are described by figure~\ref{fig3}, where we report the optimized relative signal-to-noise ratio of our detection scheme [Eq.~\rp{rmax}]
as a function of the cooperativity $\Gamma_1$ written in Eq.~\rp{Gamma1}, and the detection noise parameter $\zeta$. 
Fig.~\ref{fig3} clearly shows that efficient detection is achieved when both $\Gamma_1$ is sufficiently large and $\zeta$ sufficiently small. We emphasize that in practice the value of $\zeta$ can be reduced by enhancing the intensity of the coherent component of the reference field in the phase sensitive detection process (see appendix~\ref{App:detection}). Correspondingly, Fig.~\ref{fig4} shows that when $r_{\rm max}$ is large also the signal-to-noise ratio itself ($SNR_{\rm max}$, here reported relative to the detection time $\tau$, the probe power $\beta$  and the strength of the perturbation $\epsilon$) takes its maximum value.

These results are obtained at fixed temperature. Fig.~\ref{fig5} shows that, as expected, the signal-to-noise ratio decreases with increasing temperatures. However, it remains significantly larger than the corresponding value obtained with a simple LC resonator over a large range of temperatures.  
Apart from extremely low/unrealistic temperatures, the nonreciprocal detection scheme proposed here is more efficient, and the ratio $r_{\rm max}$ is almost constant over several orders of magnitude.
As observed in Fig.~\ref{fig6} this behavior remains almost the same regardless the exact values of $\Gamma_1$ and $\zeta$. 

The reduced efficiency at very low temperature is due to the fact that in this regime rf noise is already very low so that 
the improvement gained by its suppression is not sufficient to counterbalance the negative effect of using a more complex device with more dissipative channels.

\subsection{Effect of the phase $\varphi$ on the performance of the 4-modes model}

Finally, in Fig.~\ref{fig0.3}, we study the effect of the phase $\varphi$~\rp{varphi} on the performance of our strategy. This phase is relevant only for the 4-modes model. It is the phase difference accumulated along the two transmission paths, that connect the two electromagnetic output ports, corresponding to the two mechanical resonators
[see Fig.~\ref{fig1} (c)]. These results show that while the system is relatively sensitive to variations in this phase, strong enhanced sensitivity can be obtained by properly controlling the driving field detunings
%, which determine the value of 
$\delta$ %according to 
[Eq.~\rp{resonance}]
%, 
over a wide range of phases. Fig.~\ref{fig0.3} also shows that, in our proposal, nonreciprocity (which can be obtained by properly tuning this phase~\cite{Eshaqi}) provides no particular advantage for sensing contrary to what  suggested for similar systems in Refs.~\cite{Clerk,mcdonald2020,Nori}.

%%%%%%%%%%%%%%%%%%%%%%%%%%%%%%%%%%%%%%%%%%%%%%%%%%%%%%%%%%%%%%%%%%%%%%%%%%%
%%%%%%%%%%%%%%%%%%%%%%%%%%%%%%%%%%%%%%%%%%%%%%%%%%%%%%%%%%%%%%%%%%%%%%%%%%%%
%%%%%%%%%%%%%%%%%%%%%%%%%%%%%%%%%%%%%%%%%%%%%%%%%%%%%%%%%%%%%%%%%
\section{Conclusions}\label{conclusions}

Our investigation has focused on utilizing opto-electromechaincal systems to improve the sensitivity in detecting weak variations in a system parameter by the measurement of the system response at radio frequencies.
Here we have considered optoelectromechanical systems which include either one or two mechanical resonators, and we have discussed the possibility to
achieve enhanced sensing of a generic external parameter affecting the capacitance of the rf resonator. We demonstrated that strong enhanced sensing can be obtained close to a regime of impedance matching in which the reflection of rf-input field is strongly suppressed and the thermal fluctuations are redistributed among the system elements, resulting in a strong reduction of the rf output noise. 
Our findings show that the sensitivity in detecting the variation of the system capacitance by measuring a probe rf-field reflected by our system
is markedly enhanced compared to a similar measurement conducted with a simple rf resonator. 
The improved sensitivity is achieved under the conditions of sufficiently large optomechanical and electromechanical cooperativities and detection noise that is not excessively large.
This protocol is efficient over a wide range of temperatures. In particular it is efficient also in regimes of high thermal noise. 

\acknowledgments{We acknowledge the support of PNRR MUR project PE0000023-NQSTI (Italy).}

\appendix

\section{The small parameters $\wt g_{2j}$, $\wt\omega_{LC}$ and $\wt\Delta$}
\label{wtpars}

The parameters $\wt g_{2j}$ introduced in Eq.~\rp{gepsilon} are given by 
\begin{eqnarray}
\wt g_{21}&=&g_{21}\ v\ ,
\nn\\
\wt g_{22}&=&g_{22}\ v^*\ ,
\end{eqnarray}
where
\begin{eqnarray}\label{v}
{v}&=&
\frac{1}{4}-\frac{
\pt{\frac{\gamma_{LC}}{2}+\ii\ \omega_X}^2
}{
\omega_{LC}\al{0}{}^2+\pt{\frac{\gamma_{LC}}{2}+\ii\ \omega_X}^2
}\ .
\end{eqnarray}
Moreover the definitions of the parameters $\wt\omega_{LC}$ and $\wt\Delta$ introduced in Eqs.~\rp{omegaLCepsilon}a and \rp{Deltaepsilon} are related to the time independent part of the mechanical amplitudes [see Eq.~\rp{deltab}] 
\begin{eqnarray}
\beta_{j}\al{dc}\Bigl|_{\pp{l}
{
{C_0\ \to\ C_0(1+\epsilon)}
\\
C'_j\ \to\ C'_j(1+\epsilon)
}}&\simeq&
\beta_{j}\al{dc}
+\epsilon\ 
\pq{\frac{1}{2}+2\ {\rm Re}\pt{\, {v}}\,}
\ \beta_{j}\al{dc}\Bigl|_{g_{0,1j}\to0}
\end{eqnarray}
where
\begin{eqnarray}
\beta_{j}\al{dc}\Bigl|_{g_{0,1j}\to 0}
&=&
\frac{
i\ \abs{g_{2j}}^2
}{
2\ g_{0,2j}\pt{\frac{\gamma_m}{2}+i\,\omega_j}
}\ ,
\end{eqnarray}
such that
\begin{eqnarray}
\label{omegaLCt}
\wt\omega_{LC}
&=&
 4\
\pq{\frac{1}{2}+2\ {\rm Re}\pt{\, {v}}\,}\
\sum_{j} g_{0,2j} Re\pg{\wt\beta_{j}\al{dc}\Bigl|_{g_{0,1j}\to 0}}
\\
\label{Deltat}
\wt\Delta
&=&
2\
\pq{\frac{1}{2}+2\ {\rm Re}\pt{\, {v}}\,}\ 
\sum_{j} g_{0,1j} Re\pg{\wt\beta_{j}\al{dc}\Bigl|_{g_{0,1j}\to 0}}
\ .
\end{eqnarray}

\section{The parameters $X$ and $Y$ in Eqs.~\rp{signal} and \rp{SNRa}}
\label{XY}

The parameters $X$ and $Y$ in Eqs.~\rp{signal} and \rp{SNRa} are small and their effect can be neglected. 
They are related to the small parameters $\wt\omega_{LC}$, $\wt\Delta$ and  $\abs{\wt g_{2j}}$ introduced in Eqs.~\rp{gepsilon}-\rp{Deltaepsilon} and defined in App.~\ref{wtpars}.

In the 4-modes case they are equal to
\begin{eqnarray}\label{X}
X&=&\frac{2\ \wt\omega_{LC}}{\omega_{LC}}-
\frac{w\ \Gamma_1\ \gamma_{LC}\ \wt\Delta}{\kappa\ \omega_{LC}}\
\frac{w/\Gamma_2+\cos\varphi-1}{1+\Gamma_1\pt{1-\cos\varphi}}
\\
Y&=&\frac{2\ w\ \gamma_{LC}}{\omega_{LC}}\pq{{\rm Re}(\,{v}\,)+\frac{{\rm Im}(\,{v}\,)\ \Gamma_1\ \sin\varphi}{1+\Gamma_1\pt{1-\cos\varphi}}}
\label{Y}
\end{eqnarray}
with $w$ given in Eq.~\rp{ww}.
And in the 3-modes case
\begin{eqnarray}\label{X3}
X&=&\frac{2\ \wt\omega_{LC}}{\omega_{LC}}-
\frac{w^2\ \Gamma_1\ \gamma_{LC}\ \wt\Delta}{\Gamma_2\ \kappa\ \omega_{LC}}
\\
Y&=&\frac{2\ w\ \gamma_{LC}}{\omega_{LC}}\ {\rm Re}(\,{v}\,)\ .
\label{Y3}
\end{eqnarray}
with $w$ given in Eq.~\rp{w3}.

In both cases, the parameters $v$, $\wt\omega_{LC}$ and $\wt\Delta$ are those defined in Eqs.~\rp{v}, \rp{omegaLCt} and \rp{Deltat}.

\section{Maximization of the signal-to-noise ratio as a function of $w$}
\label{maximization}

The expression for the signal-to-noise ratio in Eq.~\rp{SNRa} can be rewritten as
\begin{eqnarray}\label{SNRapp}
SNR&=&
\frac{
16\ \abs{\beta}^2\ \epsilon^2\ \tau\ \eta\
\omega_{LC}^2/\gamma_{LC}^2\ \pq{\pt{1+X}^2+Y^2}
}{
\sigma\ 
\pt{1+w}^2\pq{
w^2-2\pt{1-2\,\xi}\ w+1
}
}
\nn\\
\end{eqnarray}
where
\begin{eqnarray}
\xi=\frac{\varrho}{\sigma}
\end{eqnarray}
with $\varrho$  and $\sigma$ defined in Eqs.~\rp{varrho} and \rp{sigma}.
In particular, 
when $\omega_{LC}\gg\gamma_{Lc}$, the parameters $X$ and $Y$ (see App.~\ref{XY}) are small and can be neglected (note that we verified the validity of this approximation in Figs.~\ref{fig0.2}, \ref{fig4} and \ref{fig5}), so we can maximize this expression by
focusing solely on minimizing the denominator of $SNR$ 
\begin{eqnarray}
d(w)=\pt{1+w}^2\pq{
w^2-2\pt{1-2\,\xi}\ w+1
}\ . 
\end{eqnarray} 
We notice that the parameter $\xi$ is always larger than zero.
Moreover, we find that $w$ is positive, and  $d(w)$ has a local minimum for positive $w$ when $\xi\leq\frac{1}{9}$.
Specifically, the local minimum is found (for $0<\xi\leq\frac{1}{9}$) when $w=w_{opt}$ with [see also Eq.~\rp{wopt}]
\begin{eqnarray}\label{Appwopt}
w_{opt}
\equiv\frac{3}{2}\pq{\frac{1}{3}-\xi+\sqrt{\pt{1-\xi}\pt{\frac{1}{9}-\xi}}
}\ .
\end{eqnarray}
This value of $w$ decreases monotonically as a function of $\xi$ and its maximum is equal to 1. This, in turn justifies the fact, that when $\omega_{LC}\gg\gamma_{Lc}$, the maximum of the signal-to-noise ratio can be approximated by neglecting the term $X$ and $Y$  in the numerator of Eq.~\rp{SNRapp}, even if they contain terms proportional to $w$ and $w^2$ (see App.~\ref{XY}).
The minimum value of the denominator, corresponding to Eq.~\rp{Appwopt}, is
\begin{eqnarray}
d(w_{opt})=\frac{4}{3}\pt{1-\xi}^2\pg{
1-\pq{\frac{9}{4}\ 
\pt{\frac{1}{9}-\xi}\
\pt{1+\sqrt{
\frac{1-\xi}{\frac{1}{9}-\xi}
}}
}^2
}\ .
\nn\\
\end{eqnarray}
This quantity increases monotonically with increasing $\xi$ and approaches the value zero in the limit $\xi\to 0$.
Moreover, this quantity is smaller than $d(0)=1$ (that gives the value of $SNR$ achieved with a simple LC circuit, meaning that a finite cooperativity $w>0$, allows for enhanced sensing) when 
$\xi<\ovl\xi=\frac{1}{48} \lpg{37 - 
\pq{23\times 277 - \pt{8\times 3\times 13}^{3/2} }^{1/3} 
}-\rpg{
\lpq{23\times 277
}+\rpq{ 
\pt{8\times 3\times 13}^{3/2} }^{1/3}
   }\sim 0.097<1/9$.
For larger values of $\xi$, the value of $d(w)$ is minimized for $w=0$ (i.e. for $\Gamma_2=0$), that corresponds to having no interaction between the LC and mechanical resonators, or in other terms,  when $\xi$ is larger than this value, the simple LC circuit is more efficient than our optoelectromechanical system for sensing.

\section{Phase sensitive detection, detection noise and detection efficiency}\label{App:detection}

In phase sensitive detection strategies the signal $V_{sig}(t)$ (here the rf-output) is multiplied by a reference field $V_{ref}(t)$~\cite{scofield1994}. The resulting detected field takes the form
\begin{eqnarray}
V_{PSD}=V_{sig}(t)\ V_{ref}(t)+f(t)\ ,
\end{eqnarray}
where we also included additional detection noise $f(t)$, characterized by $\av{f(\omega)}=0$ and by the power spectrum $S_f(\omega)$, such that $\av{f(\omega)\ f(\omega')}=S_f(\omega)\ \delta(\omega+\omega')$ (this noise term includes all the possible electronic noise, for example the noise in additional amplification or filtering stages).
Assuming that the reference is at the frequency $\omega_{LC}$ (that is equal to the frequency of the probe field $\beta$ (see Eq.~\rp{beta})% 
), and using the quantum mechanical description of the main text (with the operator $\hat a_j\al{out}$ defined in interaction picture with respect to Eq.~\rp{H0}) we define
\begin{eqnarray}\label{Vsig}
V_{sig}(t)&=&\hat a_j\al{out}\ \ee^{-\ii\ \omega_{LC}\ t}+\hat a_j\al{out}{}\da\ \ee^{\ii\ \omega_{LC}\ t}
\nn\\&=&
\pt{\alpha+\hat a}\ \ee^{-\ii\pt{ \omega_{LC}\ t+\phi_\alpha}}+\pt{\alpha^*+\hat a\da}\ \ee^{\ii\pt{ \omega_{LC}\ t+\phi_\alpha}}
\end{eqnarray}
where (see Eqs.~\rp{beta}, \rp{aoutSain} and \rp{a2outSaoin2})
\begin{eqnarray}\label{a2alphaa}
\alpha&=&\abs{\av{\hat a_2\al{out}}}=\abs{\beta\ \SSS_{2,2}}
\nn\\
\ee^{-\ii\,\phi_\alpha}&=&\frac{\av{\hat a_2\al{out}}}{\alpha}
=\frac{\beta\ \SSS_{2,2}}{\alpha}
\nn\\ 
\hat a&=&\pt{\hat a_2\al{out}-\av{\hat a_2\al{out}}}\ee^{\ii\,\phi_\alpha}=\ee^{\ii\,\phi_\alpha}\ \pg{\SSS\,\va_{in}}_2\Bigr|_{\beta=0}
\ ,
\end{eqnarray}
and
\begin{eqnarray}
V_{ref}(t)&=&\pt{\alpha_r+\hat a_r}\ \ee^{-\ii\ \pt{\omega_{LC}\ t+\phi}}+\pt{\alpha_r+\hat a_r\da}\ \ee^{\ii\ \pt{\omega_{LC}\ t+\phi}}
\end{eqnarray}
where
$\phi$ is an additional detection phase which can be tuned to detect any specific quadrature of the signal field.
Moreover, the fluctuation term $\hat a_r$ in the reference field is characterized by thermal noise, with $n_r(\omega)$ thermal excitations, such that 
\begin{eqnarray}
\av{a_r\da(\omega)\ a_r(\omega')}=n_r(\omega)\ \delta(\omega+\omega')\ .
\end{eqnarray}
In this way we find
\begin{eqnarray}
V_{PSD}&=&2\ \alpha\ \alpha_r\ \pq{\cos\pt{\phi-\phi_\alpha}+\cos\pt{2\omega_{LC}\,t+\phi+\phi_\alpha}}
\\&&
+\alpha\pq{\hat x_r\al{\phi_\alpha-\phi}+\hat x_r\al{-2\omega_{LC}\,t-\phi_\alpha-\phi}}
\nn\\&&
+
\alpha_r\pq{
\hat X_a\al{\phi-\phi_\alpha}+\hat X_a\al{2\omega_{LC}\,t+\phi+\phi_\alpha}
}+f
\nn
\end{eqnarray} 
with
\begin{eqnarray}
\hat x_r\al{\varphi}&=&\hat a_r\ \ee^{\ii\,\varphi}+\hat a_r\da\ \ee^{-\ii\,\varphi}
\nn\\
\hat X_a\al{\varphi}&=&\frac{\alpha_r+\hat a_r\da}{\alpha_r}\hat a\ \ee^{\ii\,\varphi}+\frac{\alpha_r+\hat a_r}{\alpha_r}\hat a\da\ \ee^{-\ii\,\varphi}\ .
\end{eqnarray}
In particular, when the amplitude $\alpha_r$ of the reference field is sufficiently strong as compared to the corresponding thermal excitations $\alpha_r^2\gg n_r$, it is valid to approximate the terms $\hat X_a\al{\varphi}$ as
\begin{eqnarray}
\hat X_a\al{\varphi}&\simeq&\hat a\ \ee^{\ii\,\varphi}+\hat a\da\ \ee^{-\ii\,\varphi}\ .
\end{eqnarray}
In addition, the high-frequency components (the terms at the frequency $2\omega_{LC}$ ) can be eliminated using a low-pass filter, so that the resulting detected signal is described by
\begin{eqnarray}
V_{PSD}\al{f}&=&2\ \alpha\ \alpha_r\ \cos\pt{\phi-\phi_\alpha}
+
\alpha_r\ 
\hat X_a\al{\phi-\phi_\alpha}+\hat F_0
\nn\\&=&
\alpha_r\ \hat{X}_{2}^{(out)}+\hat F_0
\end{eqnarray} 
where $\hat{X}_{2}^{(out)}$ is defined in Eq.~\rp{X2out} in terms of $\hat a_2\al{out}=\pt{\alpha+\hat a} \ee^{-\ii\,\phi_\alpha}$ (see also Eqs.~\rp{Vsig} and \rp{a2alphaa}), and $\hat F_0$ describes the total noise
\begin{eqnarray}
\hat F_0&=&\alpha\ \hat x_r\al{\phi_\alpha-\phi}
+f\ ,
\end{eqnarray}
with noise spectral density $S_F(\omega)$, such that  $\av{\hat F_0(\omega)\ \hat F_0(\omega')}=S_F(\omega)\ \delta(\omega+\omega')$, which is explicitly given by
\begin{eqnarray}
S_F(\omega)&=&{\alpha}^2\ \pq{1+2\,n_r(\omega)}+S_f(\omega)\ .
\end{eqnarray}
Now we normalize $V_{PSD}\al{f}$ by multiplying it by a factor $N$  such that the resulting quantity $\wt V=N\ V_{PSD}\al{f}$ can be rewritten in the form (see Eq.~\rp{hatm})
\begin{eqnarray}
\wt V&=&
\sqrt{\eta}\ \hat{X}_{2}^{(out)}+\sqrt{1-\eta}\ \hat F\al{noise}
\end{eqnarray}
where
\begin{eqnarray}
\eta&=&\pt{ N\ \alpha_r   }^2
\nn\\
\hat F\al{noise}&=&\frac{N\ \hat F_0}{\sqrt{1-\eta}}\ ,
\end{eqnarray}
and, correspondingly, we find that 
\begin{eqnarray}
\av{\hat F\al{noise}(\omega)\ \hat F\al{noise}(\omega')}&=&
S_{noise}(\omega)\ \delta(\omega+\omega')
\end{eqnarray}
with
\begin{eqnarray}
S_{noise}(\omega)&=&
\frac{N^2\ S_F(\omega)}{1-\eta}
\nn\\
&=&
\frac{N^2\pq{\alpha^2\ \pq{1+2\,n_r(\omega)}+S_f(\omega)} }{1-\pt{ N\ \alpha_r   }^2}\ .
\end{eqnarray}
This quantity is equal to one at zero frequency, $S_{noise}(0)=1$ (note that according to the interaction picture defined in Eq.~\rp{H0}, $\omega=0$ corresponds to the resonant frequency of the rf-resonator), when
\begin{eqnarray}
N^2=\frac{1}{\alpha_r^2+ \alpha^2\ \pq{1+2\,n_r(0)}+S_f(0)}\ .
\end{eqnarray} 
In this case, the detection efficiency $\eta$ is defined as
\begin{eqnarray}\label{eta0}
\eta=\frac{\alpha_r^2}{\alpha_r^2+ \alpha^2\ \pq{1+2\,n_r(0)}+S_f(0)}\ .
\end{eqnarray}
In particular, assuming that the detection noise $f$ and the thermal noise of the reference field are white (i.e. $S_{f}(\omega)$ and $n_r(\omega)$ are frequency independent) on the relevant range of frequency of our device (a range of frequency $\Delta\omega>>\gamma_{LC}$ around the resonance frequency $\omega_{LC}$), then we find the model of Sec.~\ref{measurement}.

Finally, 
we note that for any realistic temperature, at rf frequencies
\begin{eqnarray}
1+2\,n_r\simeq 2\,\frac{k_B\ T}{\hbar\ \omega_{LC}}
\end{eqnarray}
moreover,
if we further assume that $S_f$ is essentially Johnson–Nyquist noise, which is proportional to the temperature, such that 
$$S_f\equiv \ovl s_f\ T\ ,$$ 
then
we introduce the parameter 
\begin{eqnarray}
\zeta\equiv\frac{1}{\alpha_r^2}\pt{\frac{2\ k_B}{\hbar\ \omega_{LC}}+\ovl s_f}
\end{eqnarray}
and we can rewrite Eq.~\rp{eta0} as the Eq.~\rp{etaZeta} of the main text
where $\zeta\, T$ describes the detection noise power relative to the intensity of the coherent component of the reference field.


\begin{thebibliography}{99}


\bibitem{Degen}
C. L. Degen, F. Reinhard, and P. Cappellaro,
Quantum sensing,
\href{https://journals.aps.org/rmp/abstract/10.1103/RevModPhys.89.035002}
{Rev. Mod. Phys., {\bf 89}, 035002 (2017).} 

\bibitem{pirandola2018}  S. Pirandola, B. R. Bardhan, T. Gehring, C. Weedbrook, S. Lloyd,
 Advances in photonic quantum sensing,
  \href{https://doi.org/10.1038/s41566-018-0301-6}{Nature Photonics, {\bf 12}, 724 (2018)}.

%%%%%%%%%%%%%%%%%%%%%%%%%%%%%%%%%%%%%%%%


\bibitem{bagci2014}  
T. Bagci, A. Simonsen, S. Schmid, L. G. Villanueva, E. Zeuthen, J. Appel, J. M. Taylor, A. Sørensen, K. Usami, A. Schliesser, and E. S. Polzik, Optical detection of radio waves through a nanomechanical transducer, \href{https://doi.org/10.1038/nature13029}{Nature, {\bf 507}, 81 (2014)}.
%
\bibitem{moaddelhaghighi2018}  
I. Moaddel Haghighi, N. Malossi, R. Natali, G. Di Giuseppe, D. Vitali, Sensitivity-Bandwidth Limit in a Multimode Optoelectromechanical Transducer, \href{https://doi.org/10.1103/PhysRevApplied.9.034031}{Phys. Rev. Applied, {\bf 9}, 034031 (2018)}.
%
\bibitem{takeda2018b}  
K. Takeda, K. Nagasaka, A. Noguchi, R. Yamazaki, Y. Nakamura, E. Iwase, Jacob M. Taylor, and Koji Usami, Electro-mechano-optical detection of nuclear magnetic resonance, \href{https://doi.org/10.1364/OPTICA.5.000152}{Optica, {\bf 5}, 152–158 (2018)}.
%
\bibitem{simonsen2019b}  
A. Simonsen, J. D. Sánchez-Heredia, S. A. Saarinen, J. H. Ardenkjær-Larsen, A. Schliesser, and E. S. Polzik, Magnetic resonance imaging with optical preamplification and detection, \href{https://doi.org/10.1038/s41598-019-54200-3}{Sci. Rep., {\bf 9}, 18173 (2019)}.
%
\bibitem{Malossi}
N. Malossi, P. Piergentili, J. Li, E. Serra, R. Natali, G. Di Giuseppe, and D. Vitali,
Sympathetic cooling of a radio-frequency LC circuit to its ground state in an optoelectromechanical system,
\href{https://doi.org/10.1103/PhysRevA.103.033516}
{Phys. Rev. A, {\bf 103}, 033516 (2021)}.
%
\bibitem{Jiang}
Z. Jiang, H. Cai, R. Cernansky, X. Liu, and W. Gao, 
Quantum sensing of radio-frequency signal with NV centers in SiC,
\href{https://www.science.org/doi/10.1126/sciadv.adg2080}
{Sci. Adv., {\bf 9}, eadg2080 (2023).}

%
\bibitem{Bonaldi}M. Bonaldi, A. Borrielli, G. Di Giuseppe, N. Malossi, B. Morana, R. Natali, P. Piergentili, P.M. Sarro, E. Serra, D. Vitali, Low Noise Opto-Electro-Mechanical Modulator for RF-to-Optical Transduction in Quantum Communications, 
\href{https://doi.org/10.3390/e25071087}
{Entropy, {\bf 25}, 1087 (2023)} 
%
%%%%%%%%%%%%%%%%%%%%%%%%%%%%%%%%%%%%%
\bibitem{Eshaqi}
N. Eshaqi-Sani, S. Zippilli, and D. Vitali,
Nonreciprocal conversion between radio-frequency and optical photons with an optoelectromechanical system,
\href{https://journals.aps.org/pra/abstract/10.1103/PhysRevA.106.032606}
{Phys. Rev. A, {\bf 106}, 032606 (2022).}
%%%%%%%%%%%%%%%%%%%%%%%%%%%%%%%%%%%%%
\bibitem{barzanjeh2018}
  S. Barzanjeh, M. Aquilina, and A. Xuereb,
   Manipulating the Flow of Thermal Noise in Quantum Devices, 
   \href{https://doi.org/10.1103/PhysRevLett.120.060601}{Phys. Rev. Lett., {\bf 120}, 060601 (2018)}.
%
\bibitem{MetelmannEnt}
L. Orr, Saeed A. Khan, N. Buchholz, S. Kotler, and A. Metelmann,
High-Purity Entanglement of Hot Propagating Modes Using Nonreciprocity,
\href{https://journals.aps.org/prxquantum/abstract/10.1103/PRXQuantum.4.020344}
{PRX Quantum {\bf 4}, 020344 (2023).}
%
\bibitem{tang2023}  
Zhi-Xiang Tang, and Xun-Wei Xu, Thermal-Noise Cancellation for Optomechanically Induced Nonreciprocity in a Whispering-Gallery-Mode Microresonator, \href{https://doi.org/10.1103/PhysRevApplied.19.034093}{Phys. Rev. Appl., {\bf 19}, 034093 (2023)}.
%%%%%%%%%%%%%%%%%%%%%%%%%%%%%%%%%%%%%%%%%%%%



%%%%%%%%%%%%%%%%%%%%%%%%%%%%%%%%%%%%%%%%%%%%
%%%%%%%%%%%%%%%%%%%%%%%%%%%%%%%%%%%%%%%%%
\bibitem{Clerk1}
A. Metelmann, and A. Clerk,
Nonreciprocal Photon Transmission and Amplification via Reservoir Engineering,
\href{https://doi.org/10.1103/PhysRevX.5.021025}
{Phys. Rev. X, {\bf 5}, 021025 (2015)}.

\bibitem{Xu}
Xun-Wei Xu, Yong Li, Ai-Xi Chen, and Yu-xi Liu,
Nonreciprocal conversion between microwave and optical photons in electro-optomechanical systems,
\href{https://doi.org/10.1103/PhysRevA.93.023827}
{Phys. Rev. A, {\bf 93}, 023827 (2016)}.

\bibitem{Tian}
Lin Tian, and Zhen Li,
Nonreciprocal quantum-state conversion between microwave and optical photons,
\href{https://doi.org/10.1103/PhysRevA.96.013808}
{Phys. Rev. A, {\bf 96}, 013808 (2017)}.

\bibitem{Bernier}
N. R. Bernier, L. D. T\'oth, A. Koottandavida, M. A. Ioannou, D. Malz, A. Nunnenkamp, A. K. Feofanov, and T. J. Kippenberg,
Nonreciprocal reconfigurable microwave optomechanical circuit,
\href{https://doi.org/10.1038/s41467-017-00447-1}
{Nat. Commun., {\bf 8}, 604 (2017)}.

\bibitem{Malz}
D. Malz, L. D. T\'oth, N. R. Bernier, A. K. Feofanov, T. J. Kippenberg, and A. Nunnenkamp,
Quantum-Limited Directional Amplifiers with Optomechanics,
\href{https://doi.org/10.1103/PhysRevLett.120.023601}
{Phys. Rev. Lett., {\bf 120}, 023601 (2018)}.

\bibitem{Barzanjeh}
S. Barzanjeh, M. Wulf, M. Peruzzo, M. Kalaee, P. B. Dieterle, O. Painter, and J. M. Fink,
Mechanical on-chip microwave circulator,
\href{https://doi.org/10.1038/s41467-017-01304-x}
{Nature Communications, {\bf 8}, 953 (2017)}.

%%%%%%%%%%%%%%%%%%%%%%%%%%%%%%%%%%%%%

\bibitem{Clerk}
H.-K. Lau, and A. Clerk,
Fundamental limits and nonreciprocal approaches in non-Hermitian quantum sensing,
\href{https://www.nature.com/articles/s41467-018-06477-7}
{Nature Communications, {\bf 9}, 4320 (2018).}

\bibitem{mcdonald2020}  Alexander McDonald, Aashish A. Clerk, Exponentially-enhanced quantum sensing with non-Hermitian lattice dynamics, \href{https://doi.org/10.1038/s41467-020-19090-4}{Nat. Commun., {\bf 11}, 5382 (2020)}.

\bibitem{Nori}
L. Bao, B. Qi, D. Dong, and F. Nori,
Fundamental limits for reciprocal and nonreciprocal non-Hermitian quantum sensing,
\href{https://journals.aps.org/pra/abstract/10.1103/PhysRevA.103.042418}
{Phys. Rev. A, {\bf 103}, 042418 (2021).}
%%%%%%%%%%%%%%%%%%%%%%%%%%%%%%%%%%%%%%%%%%%

%%%%%%%%%%%%%%%%%%%%%%%%%%%%%%%%%%%%%%%%%%%%%%%%%%%%%%%%%%%%%%%%%%%%%%%
\bibitem{Zippilli}
S. Zippilli, N. Kralj, M. Rossi, G. D. Giuseppe, and D. Vitali,
Cavity optomechanics with feedback-controlled in-loop light,
\href{https://journals.aps.org/pra/abstract/10.1103/PhysRevA.98.023828}
{Phys. Rev. A, {\bf 98}, 023828 (2018).}

\bibitem{scofield1994} 
John H. Scofield, Frequency-domain description of a lock-in amplifier, \href{https://doi.org/10.1119/1.17629}{American Journal of Physics, {\bf 62}, 129-133 (1994)}.
%%%%%%%%%%%%%%%%%%%%%%%%%%%%%%%%%%%%%%%%%%%%%%%%%%%%%%%%%%%%%%%%%%%%%%%%

\end{thebibliography}
\end{document}